# Sub-picosecond photon-efficient 3D imaging using single-photon sensors

Felix Heide, Steven Diamond, David B. Lindell, Gordon Wetzstein

*Stanford University, Department of Electrical Engineering*

**Active 3D imaging systems have broad applications across disciplines, including biological imaging, remote sensing and robotics. Applications in these domains require fast acquisition times, high timing resolution, and high detection sensitivity. Single-photon avalanche diodes (SPADs) have emerged as one of the most promising detector technologies to achieve all of these requirements. However, these detectors are plagued by measurement distortions known as pileup, which fundamentally limit their precision. In this work, we develop a probabilistic image formation model that accurately models pileup. We devise inverse methods to efficiently and robustly estimate scene depth and reflectance from recorded photon counts using the proposed model along with statistical priors. With this algorithm, we not only demonstrate improvements to timing accuracy by more than an order of magnitude compared to the state-of-the-art, but this approach is also the first to facilitate sub-picosecond-accurate, photon-efficient 3D imaging in practical scenarios where widely-varying photon counts are observed.**

Active 3D imaging has broad applications across disciplines, including remote sensing, non-line-of-sight imaging, autonomous driving, robotics, and microscopic imaging of biological samples [1–11]. Key requirements for most of these applications include high timing accuracy, fast acquisition rates, wide dynamic operating ranges, and high detection sensitivity. In particular, remote sensing and automotive applications [1–4,6] demand acquisition ranges from $< 1$ meter to kilometers, while non-line-of-sight imaging [7–9], for example, relies on the information encoded by the few returning photons of multiply scattered indirect light, in addition to the directly reflected light. To facilitate these applications, ultra-sensitive detectors have been developed that allow for individual photons returning from a pulsed illumination source to be recorded. The most sensitive time-resolved detectors to date are single-photon avalanche diodes (SPAD), which have been rapidly established as a core detector technology for 3D imaging [2,12–18].

SPADs are reverse-biased photodiodes operated above their breakdown voltage, i.e. in Geiger mode [12]. Photons incident on the active surface of the SPAD can trigger an electron avalanche, which is subsequently time-stamped. By repeatedly timestamping photons returning from a synchronously pulsed illumination source, typically operating at MHz rates, a histogram of photon counts over time can be accumulated. The histogram approximates the intensity of the returning light pulse and characterizes the propagation delay, enabling distance, reflectance and 3D geometry to be recovered. While a SPAD can be operated in a free-running mode [19], which allows photon events to be detected at all arrival times simultaneously, some modes of operation enable gated detection wherein only photons within a specified time window between pulses are detected [20]. In this gated mode [21,22], they achieve high accuracy and robustness to ambient light by electronically gating out all but a narrow temporal slice [20]. However, gated acquisition requires a sequential sweep of the temporal slice over the full target range, which restricts it to applications that do not require fast acquisition rates. SPAD detectors operated in free-running mode do not suffer from this limitation, but whether these sensors are used for 3D imaging, microscopy, non-line-of-sight imaging, or any other application, they are subject to a fundamental phenomenon which severely limits accuracy: pileup distortions [23,24].

Pileup is a limitation inherent to the SPAD detector's working principle. After every triggered electron avalanche, the detector needs to be quenched before it is able to detect further photon arrival events. During this dead time, which is typically in the order of tens to hundreds of ns, the detector is inactive. Thus, earlier photons of a single laser pulse are more likely to trigger an avalanche while later ones are more likely to fall into the dead time, thus being ignored. This causes severe measurement skew, which is known as pileup. Pileup cannot be corrected in hardware and results in round-trip time-stamping errors that are orders of magnitudes larger than the timing precision of the detector.

To avoid pileup distortion, active imaging systems can be operated in a low-flux regime, where the average number of incident photons per measurement is extremely low (i.e., $\ll 1$). State-of-the-art depth and reflectivity estimation techniques have demonstrated successful operation in these challenging conditions using probabilistic image formation models and statistical priors in the inverse methods [18,25,26]. However, many 3D imaging applications, for example in robotics, biological imaging, or automotive sensing, operate in environments where both objects returning a low number of signal photons and objects reflecting higher numbers of photons are essential for decision making. The large variance in acquired photon counts of objects at different depths or resulting from varying reflectivity of different objects makes it crucial for a practical 3D imaging technique to operate robustly in both low-flux and high-flux conditions simultaneously (see Fig. 1).

In this work, we introduce a new estimation algorithm that overcomes existing limitations of active 3D imaging systems with free-running SPADs. Specifically, we demonstrate sub-picosecond timing accuracy for pulsed light sources with a full width at half maximum (FWHM) wider than 50 ps. The proposed method improves the accuracy of existing depth and albedo estimation algorithms by more than an order of magnitude across a wide dynamic range, from low-flux to high-flux measurements. These benefits are enabled by an image formation model and corresponding inverse method that lift assumptions of low-flux models [18,25] to broad operating conditions that are distorted by pileup. To this end, we introduce a probabilistic image formation model that includes pileup and we derive efficient inverse methods for the depth and albedo estimation problem with this model. The inverse methods exploit statistical priors of both depth and albedo. Unlike previous work, our reconstruction framework jointly estimates all unknown parameters, overcoming algorithmic limitations that restrict the timing precision of existing methods. The proposed methods not only facilitate highly accurate and fast 3D imaging but they also open up a new operating regime of photon-efficient 3D imaging in conditions with drastically-varying photon counts.

## Results

**Imaging setup.** The experimental acquisition setup is illustrated in Fig. 1. The light source is a pulsed, collimated laser that is scanned horizontally and vertically using a 2-axis mirror galvanometer. Light scattered back from the scene is focused on the SPAD (Micro Photon Devices PD-100-CTD) using a microscope objective. A PicoQuant PicoHarp 300 module is used for time-stamping, histogram accumulation, and synchronization of laser and detector. The dead time of the SPAD detector is 75 ns, corresponding to a path length of 22.5 m. Due to the fact that this dead time is multiple orders of magnitude larger than the desired timing accuracy, the pileup distortions of measured histograms can only be ignored in the ultra-low-flux regime with $< 10^{-2}$ average photons detected per pulse.

**Observation model.** We discretize the scene into a grid of $m \times n$ points at distances $\boldsymbol{z} \in \mathbf{R}^{m \times n}$ and denote the reflectance of these points as $\boldsymbol{\alpha} \in [0,1]^{m \times n}$. Each point is probed with $N$ pulses.

We first describe the observation model for a single point $(i, j)$ by modeling the temporal shape of the laser pulse $g$ as a mixture of Gaussians

$$g(t) = \sum_{k=1}^{K} a_k \exp\left(-\frac{(t-b_k)^2}{c_k^2}\right), \tag{1}$$

where the parameters $a_i, b_i, c_i \in \mathbf{R}$ are calibrated in a pre-processing step (cf. Methods). Given this parametric impulse model, the photon detections on the SPAD detector can be modeled as an inhomogeneous Poisson process with rate function

$$r_{ij}(t, \alpha_{ij}, z_{ij}, s) = \mu\alpha_{ij}\, g(t - 2z_{ij}/c) + s, \tag{2}$$

where $c$ is the speed of light, $\mu$ is the SPAD photon detection probability, and $s$ is background detections from ambient light and dark count [18].

The free-running SPAD records photon detections over a time interval $[0, T]$ discretized into uniform bins of size $\Delta$, where $T$ is the inverse of the laser repetition rate. Let $h_{ijk}$ denote the accumulated counts in bin $k$ over the $N$ pulses and $\lambda_{ijk}(\alpha_{ij}, z_{ij}, s) = \int_{k\Delta}^{(k+1)\Delta} r_{ij}(t, \alpha_{ij}, z_{ij}, s)dt$ the Poisson rate for the number of detections in each bin $k$. The proposed probabilistic reconstruction method infers the latent variables $\boldsymbol{z}, \boldsymbol{\alpha}$ using maximum-a-posteriori (MAP) estimation, which relies on the probability $P(\mathbf{h}_{ij}|\boldsymbol{\lambda}_{ij})$ of observing a histogram $\mathbf{h}_{ij} \in \mathbf{Z}^T$ given means $\boldsymbol{\lambda}_{ij}(\alpha_{ij}, z_{ij}, s) \in \mathbf{R}^T$ and priors on depth and albedo.

Prior work on 3D imaging using SPAD detectors focuses on the low-flux regime[18,25,27,28] in which the expected number of photon detections per pulse is significantly smaller than one. In the low-flux regime we may neglect dead time and approximate the observations $h_{ijk}$ as being conditionally independent across bins $k$, with corresponding Poisson probability mass function $P(h_{ijk}|\boldsymbol{\lambda}_{ij})$. For medium and high-flux measurements the conditional independence approximation breaks down because dead time ensures that at most one photon detection is recorded per pulse [23]. The probability of a single histogram bin, taking dead time into account, is given by the multinomial distribution

$$P(\mathbf{h}_{ij}|\boldsymbol{\lambda}_{ij}) = \frac{N!\exp(-\mathbf{1}^T\boldsymbol{\lambda}_{ij})^{N-\mathbf{1}^T\mathbf{h}_{ij}}}{h_{ij1}!\cdots h_{ijT}!(N-\mathbf{1}^T\mathbf{h}_{ij})!}\prod_{\ell=1}^{T}\left(\exp\left(-\sum_{k=1}^{\ell-1}\lambda_{ijk}\right)-\exp\left(-\sum_{k=1}^{\ell}\lambda_{ijk}\right)\right)^{h_{ij\ell}}, \tag{3}$$

where $\mathbf{1}$ is the vector of all ones. We refer to the Supplemental Methods for a detailed derivation of this probabilistic model. The observation model holds across the full measurement range, from low-flux to high-flux regimes.

**Reconstruction algorithm.** After scanning a scene, temporal histograms $\mathbf{h}_{ij}$ are available for each point $(i,j)$. To reconstruct scene reflectance and depth from these histograms, we find the MAP estimate of $\boldsymbol{\alpha}$ and $\boldsymbol{z}$, along with the ambient term $s$, using the observation model and the prior assumption that the gradients of reflectivity and depth maps are sparse. Inspired by prior work on depth and reflectivity estimation [18,25,26], we place transverse anisotropic total variation (TV) priors directly on $\boldsymbol{\alpha}$ and $\boldsymbol{z}$. The MAP estimate is given by solving the optimization problem

$$\underset{\boldsymbol{\alpha},\boldsymbol{z},s}{\text{minimize}} \quad \textstyle\sum_{i=1}^{m}\sum_{j=1}^{n} -\log P(\mathbf{h}_{ij}|\boldsymbol{\lambda}_{ij}(\alpha_{ij},z_{ij},s)) + \gamma_1\|\nabla\boldsymbol{\alpha}\|_1 + \gamma_2\|\nabla\boldsymbol{z}\|_1, \tag{4}$$

where $\boldsymbol{\alpha}$ and $\boldsymbol{z}$ are the unknown variables, and $\nabla$ is the gradient operator. To solve this problem, we develop a proximal algorithm for this non-convex optimization problem that decouples the sparsity-promoting prior terms from the likelihood term. Specifically, we minimize the joint objective by introducing slack variables $\mathbf{v}^{\boldsymbol{\alpha}}, \mathbf{v}^{\boldsymbol{z}}, \mathbf{v}^{s}$ for albedo, depth and ambient terms. We then optimize each unknown term in an alternating fashion (see Supplemental Material for details). As a result of this proximal optimization scheme, the log-likelihood minimization becomes separable in the measurement $ij$ as

$$\underset{\boldsymbol{\alpha},\boldsymbol{z},s}{\text{minimize}} \quad -\log P(\mathbf{h}_{ij}|\boldsymbol{\lambda}_{ij}(\boldsymbol{\alpha},\boldsymbol{z},s)) + \frac{\xi}{2}(\boldsymbol{z}-\mathbf{v}^{\boldsymbol{z}})^2 + \frac{\xi}{2}(\boldsymbol{\alpha}-\mathbf{v}^{\boldsymbol{\alpha}})^2 + \frac{\xi}{2}(s-\mathbf{v}^{s})^2, \tag{5}$$

which is an optimization problem over a non-linear tri-variate loss function and quadratic proximal closeness terms with scalar weight $\xi$. We solve the minimization problem with per-pixel parallel Newton Methods, as derived in the Supplemental Methods.

This reconstruction method improves prior SPAD depth imaging techniques both by accounting for dead time in the observation model, which is crucial for high and medium-flux scenarios, and by jointly estimating $\boldsymbol{\alpha}$ and $\boldsymbol{z}$ directly from the raw histogram data. Previous approaches, on the other hand, apply a sequence of transformations to the data, estimating $\boldsymbol{\alpha}$ and $\boldsymbol{z}$ in separate stages [18,25,27,28], which limits the reconstruction performance.

**Experimental validation.** We evaluate the proposed method on measurements experimentally acquired with the setup illustrated in Fig. 1. In Fig. 2, we assess the performance of the proposed method on two scenes with highly varying reflectance and depth profiles. Both scenes contain objects with complex geometries and varying reflectance properties, including specular behavior for the "Statue of David" scene and Lambertian reflectance with spatially varying albedo in the "Bas-relief" scene. For both scenes, we capture a ground truth reference measurement with a 5% Neutral Density filter in the laser path which eliminates pileup distortions by damping the source intensity. To minimize shot noise fluctuations at the low count rates, we acquire very long sequences of 6 s length per spot at 4 MHz laser repetition rate. We scan every scene at a spatial resolution of $150 \times 150$ points. Hence, a full reference measurement is acquired in $150 \cdot 150 \cdot 6$ s $= 37.5$ h per scene. The ground truth depth is extracted from long-exposure measurements using log-matched filtering [28] with the impulse response calibrated using a planar target captured under the same acquisition settings, i.e. unaffected by pileup. The experimental measurements, which serves as input for the proposed method, are acquired without any filters in the optical path.

For each scene in Fig. 2, depth and albedo reconstructions along with the corresponding error maps are shown. We show perspective renderings of the point cloud reconstructions for two different viewpoints. These results verify that the proposed method achieves high-quality depth and albedo reconstructions unaffected by scene-dependent pileup or shot-noise distortions. Specifically, we compare our approach against the conventional log-matched filter estimate [28] as well as Coates' pileup correction method [23] followed by a Gaussian fit. The effect of pileup becomes apparent in the error map for the log-matched filter estimate resulting in more than 5 mm depth error for the "Statue of David" scene. Existing methods do not effectively suppress pileup and therefore suffer from strongly scene-dependent depth precision. In contrast, our method achieves sub-picosecond accuracy independent of scene depth and reflectance, despite the FWHM of the laser pulses being longer than 50 ps. We include additional comparisons in the Supplemental Results where we also evaluate the proposed approach in the low-flux regime and demonstrate that our probabilistic method achieves close to an order of magnitude increased temporal resolution.

In Table 3, we validate the sub-picosecond accuracy of the proposed approach without using spatial priors, i.e. relying solely on the probabilistic pileup model for individual histogram measurements. Specifically, we acquire a sequence of single-point measurements of a planar Lambertian target which is moved along the optical axis using a linear motion stage for 100 measurement points uniformly spaced with 0.5 mm distance. The setups for this measurement scenario and for an additional horizontal scanline measurement are shown in the Supplemental Information. In addition to the conventional log-matched filter estimate [28] and Coates' method [23], we compare the proposed method against and Shin et al.'s method [18] applied on the Coates-corrected histogram data, which adds censoring and background signal suppression [18] to Coates' method. Table 3 shows the average absolute error in depth and round-trip time for both of these measurements. The results demonstrate that, even without spatial priors, the proposed probabilistic method outperforms existing state-of-the-art approaches by more than an order of magnitude. Adding the proposed spatial prior reduces temporal error by a factor of $2\times$ on average. We refer to the Supplemental Results for additional experimental results and extensive evaluation in simulation.

**Optimal regime of photon counts.** Next, we analyze the performance of the proposed method with a fixed dwell or exposure time for a varying incident photon flux. The proposed probabilistic method achieves optimal accuracy in an unconventional, pileup-affected regime with around 1 photon detection per pulse (see Fig. 4). This plot is generated in simulation without the use of spatial priors for the 450 nm Alphalas LD-450-50 laser with $N = 10^4$ shots. As expected, the log-matched filtering approach performs best when the measurements are neither too noisy nor too much affected by pileup. Existing sequential pileup correction methods, such as Coates' method [23], alleviate pileup and effectively extend the range where optimal performance can be reached beyond 1 expected photon detection per pulse. The accuracy of the proposed method is identical to previous methods for the low-flux regime, because we only consider a single pixel and no spatial priors are used. As the photon count increases, our approach substantially improves upon all existing methods by up to two orders of magnitude. Optimal precision is achieved for expected photon counts around 1 per pulse, which is significantly higher than the low-flux regime in which existing methods operate. These results motivate an optimal photon flux regime for 3D imaging that is far outside the conventional low-flux regime. In the Supplemental Results, we also analyze the effect of the histogram bin width on timing accuracy in this optimal flux regime. The accuracy is consistent across a broad range of histogram bin-widths from sub-2 ps up to 80 ps, demonstrating that the proposed approach not only improves on the state-of-the-art by more than an order of magnitude in accuracy, but also reduces the timing resolution requirements on the detector side.

## Discussion

Despite laser pulse widths being larger than 50 ps FWHM, the proposed probabilistic image formation model and corresponding inverse methods achieve sub-picosecond accuracy for active 3D imaging. Moreover, the proposed methods achieve this precision across a wide dynamic range, from low-flux to high-flux measurements. These capabilities are achieved by accurately modeling pileup distortions in the image formation model and by solving the inverse problem jointly for all latent variables using statistical priors. This paves the way for fast and precise photon-efficient 3D imaging systems in practical scenarios where widely-varying photon counts are observed in a scene. Applications across disciplines may significantly benefit from these capabilities, including 3D mapping and navigation, autonomous driving, robotic and machine vision, geographic information science, industrial imaging, and microscopic imaging.

## Methods

**Equipment details.** The hardware setup consists of a time-resolved sensor, picosecond laser, illumination and collection optics, and a set of scanning mirrors to achieve a raster scan illumination pattern. The sensor is a PDM SPAD from Micro Photon Devices with a 100 $\mu$m $\times$ 100 $\mu$m sensor area, 27 ps timing jitter (measured with a 100 kHz laser at 675 nm), and 40.9 dark counts per second. Timing of photon arrivals is captured with a PicoHarp 300 Time-Correlated Single Photon Counting (TCSPC) module. The illumination source is a 450 nm or 670 nm picosecond laser (ALPHALAS PICOPOWER-LD-450-50, PICOPOWER-LD-670-50). The 450 nm and 670 nm versions have pulse widths of 90 ps and 50 ps and average power of 0.406 mW and 0.11 mW respectively at a 10 MHz pulse repetition rate. The collection optics are designed to extend the field of view of the SPAD across the area scanned by the illumination source and consist of a 75 mm objective lens (Thorlabs AC508-075-A-ML), a 30 mm relay lens (Thorlabs AC254-030-A-ML) and a microscope objective (Olympus UPLFLN 20x objective). The laser spot is minified using a 50 mm (Thorlabs AC254-050-A-ML) and 250 mm (Thorlabs AC254-250-A-ML) lens relay and scanned with mirrors driven by a two-axis galvanometer (Thorlabs GVS012).

**Calibration details.** All parameters of the image formation model are described in the Observation model section. The detector's photon detection probability $\mu = 0.34$ at 450 nm ($\mu = 0.33$ at 670 nm) and dark count rate $d = 40.9\ c/s$ were calibrated using the method of Polyakov [29]. The remaining unknowns are $\{a_k, b_k, c_k \mid k \in 1, \ldots, K\}$ for the Gaussian mixture model $\tilde{g}$. We use $K = 8$ and calibrate for the 450 nm Alphalas LD-450-50 laser the values $\{$(0.57, 220.4, 32.0), (0.0059, 203.8, 10.2), (0.003, 214.6, 18.7), (-0.57, 220.3, 32.0), (0.003, 255.6, 47.5), (0.002, 297.9, 67.4), (0.009, 199, 7.1), (0.0003, 357.4, 143.2)$\}$, and for 670 nm Alphalas LD-670-50 laser the values $\{$(0.04, 199.4, 6.4), (0.004, 206.5, 7.0), (-0.001, 187.9, 12.5), (0.005, 204.8, 17.8), (0.003, 225.1, 27.2), (0.002, 254.6, 42.3), (0.0008, 301.0, 69.3), (0.0003, 388.7, 136.0)$\}$. Measurements of a scene consisting of a single diffuse reflector placed at 1 m distance are acquired to estimate these parameters. Specifically, an ND filter with $1\%$ transmission is placed in the illumination path, and a low-noise histogram of a single point reflector is accumulated with $N = 10^9$ pulses, requiring approximately 17 minutes at a 1 MHz laser repetition rate. The high-absorption ND filter ensures a low-flux regime where pileup can be ignored and where the histogram counts are Poisson distributed. For the high number of $10^9$ shots, these histogram measurements can be well approximated by a Gaussian distribution. Finally, the mixture parameters are estimated using efficient, conventional expectation-maximization algorithms [30]. Please see the Supplemental Methods for calibration results for the for the 450 nm Alphalas LD-450-50 laser and the 670 nm Alphalas LD-670-50 laser.

**Algorithm parameters.** The algorithm hyper-parameters are $\gamma_1 = \gamma_2 = 10^{-1}$. These values were determined using simulations. Initial values of $\boldsymbol{\alpha} = 1, \boldsymbol{z} = 0$ are used for the reconstruction algorithm in the Reconstruction algorithm section. In practice, warm-starting with the log-matched filter or Coates' estimate reduces the iterations needed to converge. The termination criteria for the proposed non-convex optimization algorithm are detailed in the Supplemental Methods. The algorithm run time with unoptimized Matlab code averages around 100 secs on a Intel i7 2.6GHz notebook computer. Note that the likelihood proximal operator, which dominates the algorithm runtime, is embarrassingly parallel across all measurement points, and hence is suited for implementation on modern GPU hardware.

**Code and data availability.** The code and data used to generate the findings of this study will be made public on GitHub.

**Acknowledgements** This work was in part supported by a National Science Foundation CAREER award (IIS 1553333), by a Sloan Fellowship, by the DARPA REVEAL program, and by the KAUST Office of Sponsored Research through the Visual Computing Center CCF grant. The authors would like to thank Rafael Setra, Kai Zang, Matthew O'Toole, Amy Fritz, and Mark Horowitz for fruitful discussions in early stages of this project.

**Supplementary information** Supplementary Information is attached to this submission.

**Competing interests** The authors declare that they have no competing financial interests.

**Author contributions** F.H. and G.W. conceived the idea. F.H. derived the algorithm. F.H. and S.D. implemented the algorithm. F.H processed all data. F.H., D.B.L., and G.W. designed the experiments. D.B.L. captured experimental data. F.H., S.D., and G.W. wrote the manuscript. G.W. supervised all aspects of the project.

**Correspondence** Correspondence should be addressed to Felix Heide and Gordon Wetzstein.

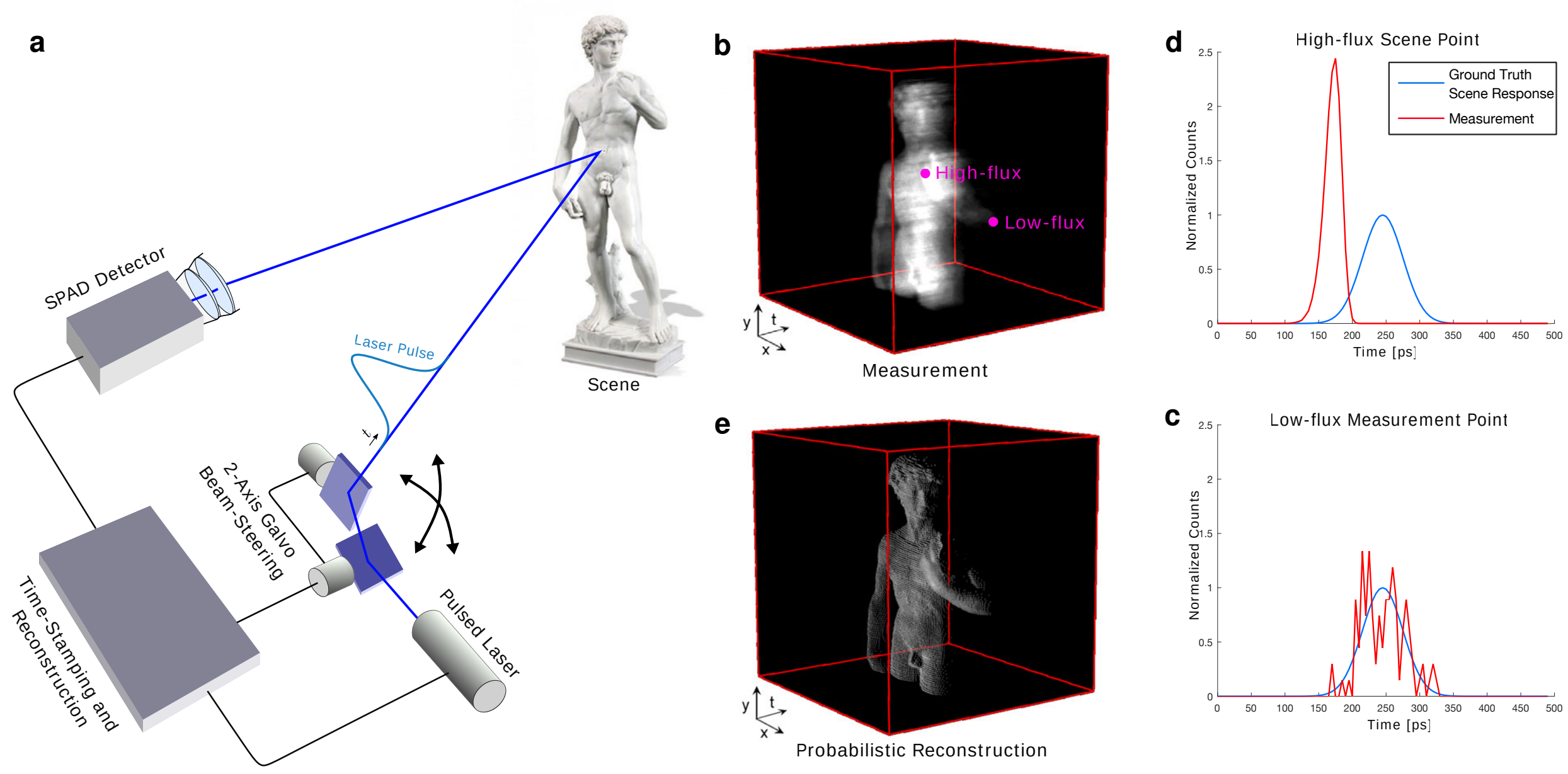

**Figure 1: Sub-picosecond 3D Imaging Framework.** **(a)** A collimated, pulsed laser illuminates the scene at a single point. The laser is laterally scanned using a 2-axis mirror galvanometer. Timing and control electronics time-stamp each detected photon arrival relative to the last emitted pulse and accumulate these events in a histogram of spatio-temporal photon counts **(b)**. This histogram is processed to estimate both reflectivity and depth information **(c)**. Two points are highlighted, one corresponding to high-flux **(d)** and the other to low-flux **(e)** measurements. Whereas the latter are noisy, high-flux measurements suffer from pileup distortion which introduce a significant bias for the depth estimation of conventional algorithms. The proposed estimation method accurately models both of these scenarios, allowing for reflectance information and travel time to be estimated with sub-picosecond accuracy from severely distorted measurements.

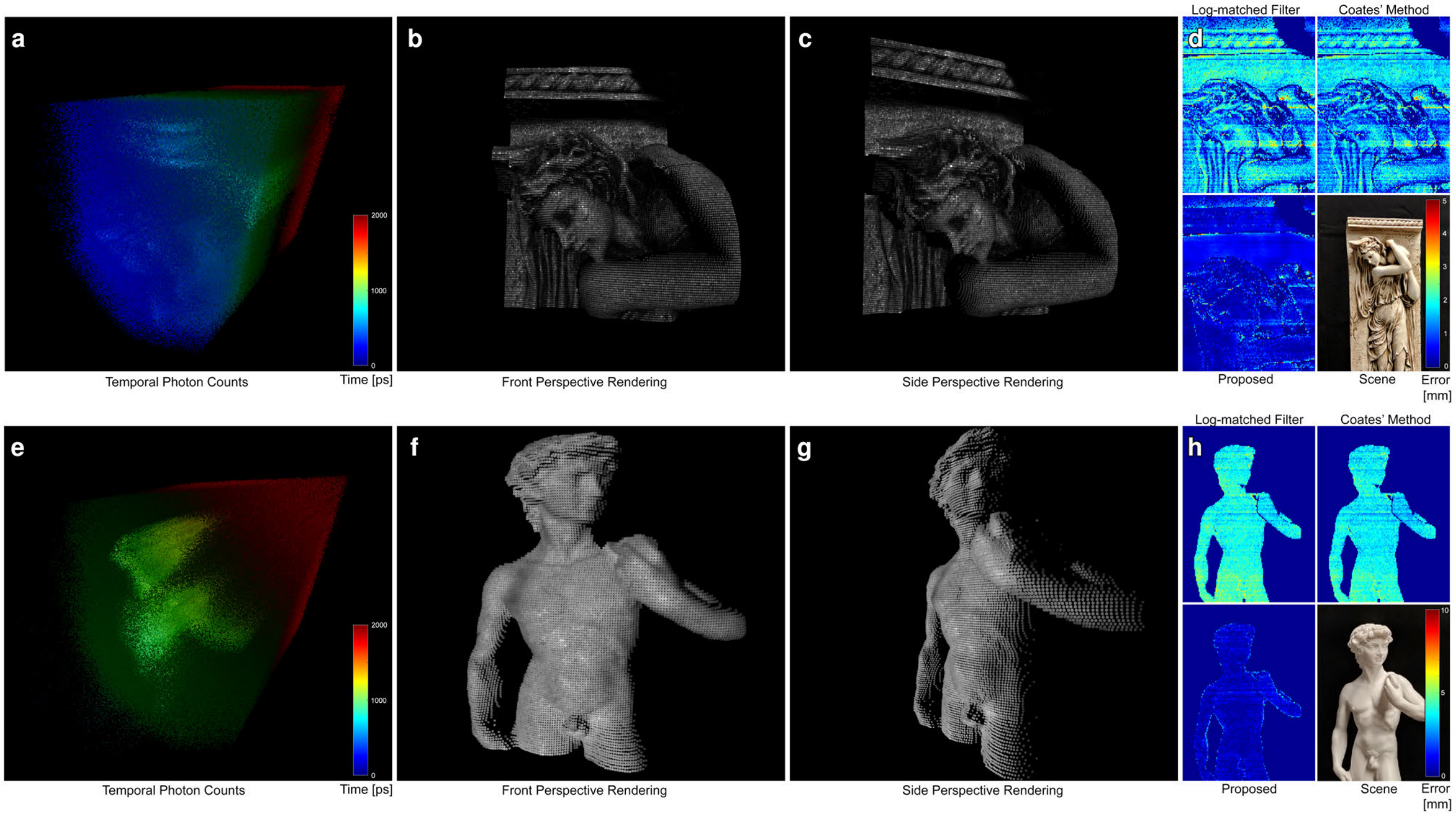

**Figure 2: Experimental reconstructions**. A recorded spatio-temporal distribution of photon counts **(a,e)** is processed to estimate a 3D point cloud **(b,c,f,g)** that contains both depth and albedo information, here shown for two different scenes (photographs shown in **(d,h)**). The color-coded errors maps **(d,h)** directly compare the results of several depth estimation techniques, including log-matched filtering [28], Coates' method [23] followed by Gaussian fit (on high-flux measurement), and the proposed method.

| Scene Target | Pulse Wavelength | Log-matched Filter Depth Error | Coates' Method [23] | First Photon [18] on Coates [23] | Proposed Method | Log-matched Filter Round-trip Error | Coates' Method [23] | First Photon [18] on Coates [23] | Proposed Method |
|---|---|---|---|---|---|---|---|---|---|
| Planar | 450 nm | 4.54 mm | 4.40 mm | 3.39 mm | **0.14 mm** | 15.13 ps | 14.68 ps | 11.30 ps | **0.46 ps** |
| Linear Stage | 450 nm | 2.79 mm | 2.65 mm | 1.98 mm | **0.24 mm** | 9.31 ps | 8.85 ps | 6.58 ps | **0.91 ps** |
| Planar | 670 nm | 3.19 mm | 3.00 mm | 2.37 mm | **0.16 mm** | 10.64 ps | 9.99 ps | 7.91 ps | **0.52 ps** |
| Linear Stage | 670 nm | 8.13 mm | 8.09 mm | 8.00 mm | **0.23 mm** | 27.11 ps | 26.97 ps | 26.68 ps | **0.76 ps** |

**Figure 3:** Experimental validation of sub-picosecond accuracy on recorded single-pixel data without spatial priors. The average depth and round-trip time error for two scenes are shown, for the 450 nm Alphalas LD-450-50 laser (FWHM of 90 ps) and the 670 nm Alphalas LD-670-50 laser (FWHM of 50 ps), respectively. The background level is $5\%$ for all scenes. We compare reconstructions of the conventional log-matched filter estimate [28], Coates' method [23] followed by a Gaussian fit, Shin et al. [18] on Coates-corrected measurements, and the proposed method.

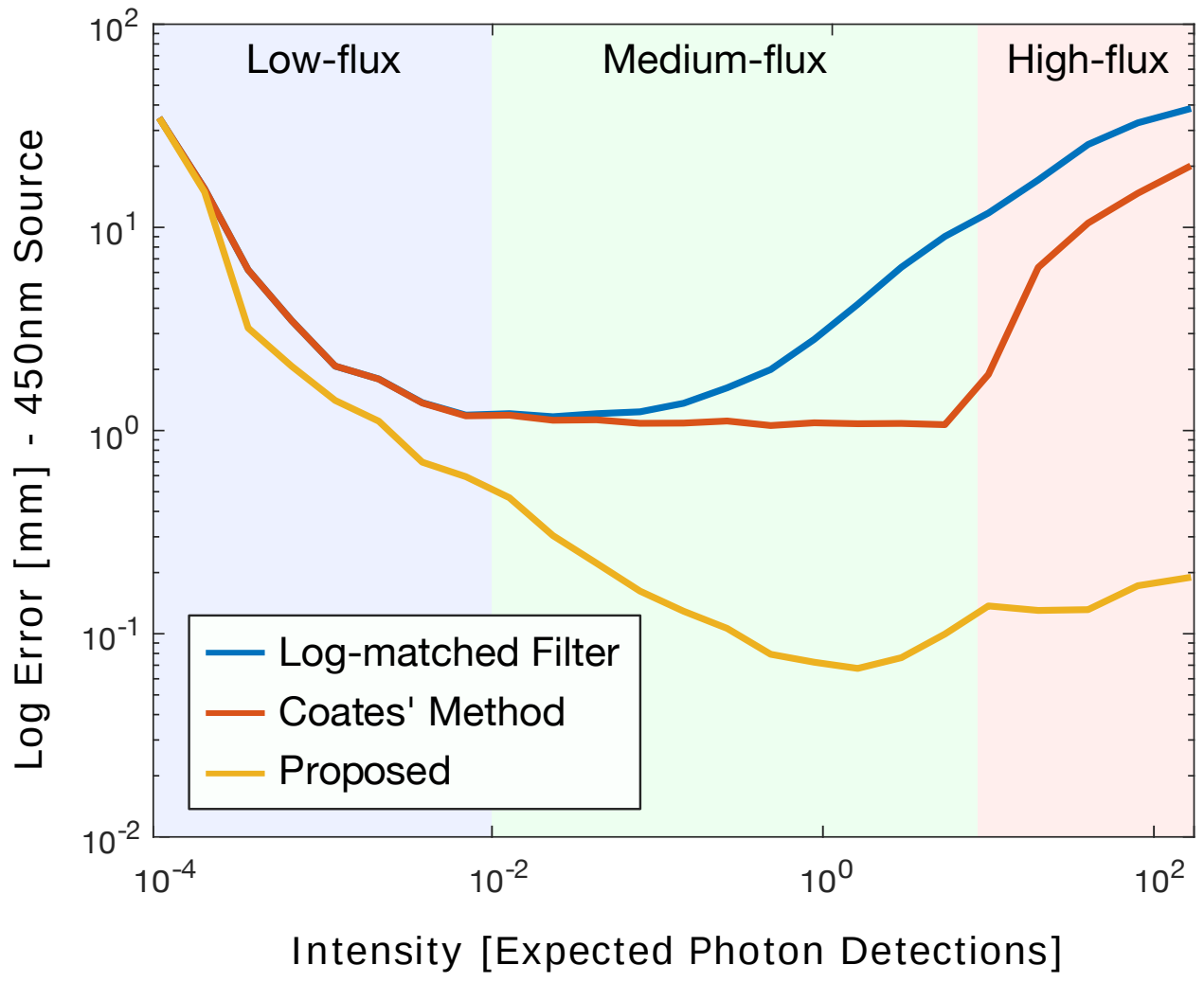

**Figure 4: Optimal photon count regime.** Depth reconstruction accuracy for varying photon counts for the 450 nm Alphalas LD-450-50 laser (FWHM of 90 ps). The conventional log-matched filter, Coates' method [23], and the proposed method are compared. The optimal number of photon counts lies around the unconventional region of 1 photon detected per pulse on average, independent of the impulse response and for a broad range of histogram bin widths, see Supplemental Results.

# Supplementary Materials and Methods

## Image Formation Model

This section introduces an observation model for imaging with single photon avalanche diodes (SPADs). For this purpose, we assume that a single SPAD, operating in free-running mode, and a pulsed laser are co-axially mounted and focused on a surface patch at some distance $z$. Additional collection optics may be mounted on the detector. The imaged surface patch is assumed to exhibit Lambertian reflectance properties.

**Illumination and Response** We illuminate the scene with $N$ laser pulses, which defines the measurement acquisition time for a given laser repetition rate. Each laser pulse is characterized by the continuous temporal impulse response function $g(t) \in \mathbf{R}_+$, which describes the source intensity as a function of time. The temporally resolved irradiance at the detector can be modeled as

$$r(t) = \alpha\, g(t - 2z/c) + s, \tag{1}$$

with $\tilde{\alpha} \in [0, 1]$ as the reflectance of the scene patch at distance $z$ and $c$ the speed of light. The additive component $s$ models ambient illumination present in the scene. The intensity $\alpha = \frac{\tilde{\alpha}}{\pi z^2} \rho\, A_c\, A_p \, \cos \measuredangle(\mathbf{n}, \mathbf{o})$ combines all intensity loss in the light transport from the source to camera via the single scene patch, consisting of the following components: distance falloff, absorption in the detector optics $\rho$, perspective foreshortening due to the orientation of the patch with normal $\mathbf{n}$ to the optical axis $\mathbf{o}$, patch size $A_p$ and aperture size $A_c$. As estimating these parameters is not central to the proposed method, we combine these terms in the factor $\alpha \in [0, 1]$.

**Detector Model** Photons impinging on the active area of the SPAD detector generate an electron avalanche. The time of occurrence of the resulting rise in voltage is digitized using a time-to-digital converter (TDC), which is synchronized with the laser. After each electron avalanche, the detector must be reset or *quenched*. As a consequence, no new photon arrival event can be detected for a certain amount of time known as the *deadtime*. The full width at half maximum (FWHM) of the laser pulse $g(t)$ is commonly orders of magnitude smaller than this deadtime, while the time between pulses $1/f$ is usually larger than the deadtime. We therefore assume that each emitted laser pulse results in at most one detected photon event. Moreover, we assume that the SPAD is fully quenched between successive laser pulses.

Oftentimes, the timing logic quantizes detected photon arrival events into a histogram with $T$ discrete bins. The laser and timing logic are synchronized so that each bin $i$ corresponds to a consistent uniform time interval $I_i = [t_i, t_{i+1})$ relative to the beginning of the emitted pulse. Throughout an acquisition, $N$ pulses are emitted and all detected photon arrival events are combined into a single histogram $\mathbf{h} \in \mathbf{Z}^T$.

We model the probability of detecting a histogram $\mathbf{h}$ given a response $r(t)$ and $N$ laser pulses as follows. Photon detections in the SPAD are either due to the incident response $r$ or dark count $d$ (false photon detections), with the latter modeled as a constant rate over time, following Kirmani et al. [1] and Shin et al. [2]. Fluctuation in the delay between the actual time of arrival of a photon and the resulting timestamp are modeled as jitter using the probability density function (PDF) $w(t)$. With these definitions, the number of photon detections in the SPAD during each interval $I_i$ can be modeled as an inhomogeneous Poisson process, with mean

$$\begin{aligned} \lambda_i &= \int_{t_i}^{t_{i+1}} \mu\big(r(z,\alpha)\ w(t)\big) + d(t)\ dt \\ &= \mu\big(\alpha\, \tilde{g}(t_i - 2z/c) + s\big) + d, \end{aligned} \tag{2}$$

where $\mu > 0$ is the quantum efficiency and $d$ is the dark count rate per interval. The system impulse response $\tilde{g}(\tau) = \int_{\tau}^{\tau+I} (g \star w)(t) dt$ absorbs both the detector jitter as well as the laser impulse response.

**Parametric Impulse Response Model** We represent the impulse response of the measurement system using a parametric continuous model that is compact and allows for fast statistical estimation. Specifically, we model $\tilde{g}$ as a Gaussian mixture model

$$\tilde{g}(\tau) = \sum_{k=1}^{K} a_k \exp\left(-\frac{(\tau - b_k)^2}{c_k^2}\right) \tag{3}$$

with $K$ mixture components. The model coefficients $\{a_k, b_k, c_k | k \in 1, \dots, K\}$ are estimated in a precalibration step using measurements acquired over long acquisition periods to reduce the effect of noise. Specifically, we find the optimal least-squares fit of the parameterized model in Eq. 3 to the measured histogram.

**Probabilistic Measurement Model** Let $\boldsymbol{\lambda}$ denote the arrival rate for all histogram bins. We can express the probability $P(n, i|\lambda)$ of $n$ photon detections in bin $i$ as a Poisson random variable with mean $\lambda_i$ that is

$$P(n, i|\boldsymbol{\lambda}) = \exp(-\lambda_i)\frac{\lambda_i^n}{n!}. \tag{4}$$

The probability $P(\text{no detections})$ of detecting no photons during a pulse is then

$$\begin{aligned} P(\text{no detections}) &= \prod_{i=1}^{T} P(n = 0, i|\boldsymbol{\lambda}) \\ &= \prod_{i=1}^{T} \exp(-\lambda_i)\frac{\lambda_i^0}{0!} \\ &= \exp(-\mathbf{1}^T\boldsymbol{\lambda}), \end{aligned} \tag{5}$$

where $\mathbf{1}$ denotes a vector with all entries equal to one.

Upon a photon detection, the detector-timing module increments the histogram bin in which the earliest photon detection occurred or does nothing if no photons were detected.

This means that photon arrivals are more likely to be recorded in earlier bins than later bins, a histogram distortion known as pile-up [3]. Specifically, the probability $P(i|\lambda)$ of recording a photon arrival event in bin $i$ is

$$\begin{aligned} P(i|\boldsymbol{\lambda}) &= \prod_{k=1}^{i-1} P(n=0,k|\boldsymbol{\lambda})P(n>0,i|\boldsymbol{\lambda}) \\ &= \prod_{k=1}^{i-1} \exp(-\lambda_k) \sum_{n=1}^{\infty} \exp(-\lambda_i) \frac{\lambda_i^n}{n!} \\ &= \prod_{k=1}^{i-1} \exp(-\lambda_k) \exp(-\lambda_i)(\exp(\lambda_i)-1) \\ &= \exp\left(-\sum_{k=1}^{i-1} \lambda_k\right) - \exp\left(-\sum_{k=1}^{i} \lambda_k\right). \end{aligned} \tag{6}$$

In the third row, we use the Taylor series $\sum_{n=1}^{\infty} \frac{x^n}{n!} = \exp(x) - 1$. Due to the fact that the dead time of the detector is less than the time between successive pulses, we assume that each of the $N$ laser pulses is an independent trial. The probability $P(\mathbf{h}|\boldsymbol{\lambda})$ of detecting the full histogram $\mathbf{h}$ is the multinomial distribution

$$\begin{aligned} P(\mathbf{h}|\boldsymbol{\lambda}) &= \frac{N!}{h_1! \cdots h_T!(N-\mathbf{1}^T\mathbf{h})!} P(\text{no detections})^{N-\mathbf{1}^T\mathbf{h}} \prod_{i=1}^{T} P(i|\boldsymbol{\lambda})^{h_i} \qquad (7) \\ &= \frac{N!}{h_1! \cdots h_T!(N-\mathbf{1}^T\mathbf{h})!} \exp(-\mathbf{1}^T\lambda)^{N-\mathbf{1}^T\mathbf{h}} \prod_{i=1}^{T} \left(\exp\left(-\sum_{k=1}^{i-1}\lambda_k\right) - \exp\left(-\sum_{k=1}^{i}\lambda_k\right)\right)^{h_i}, \end{aligned}$$

which follows from Eqs. (5) and (6).

**Low-Flux Regime** In a low-flux regime, where the probability of observing multiple photons reflected back to the detector from a single emitted pulse is small, the effect of pile-up is negligible and can be ignored [1]. In this scenario, where the average number of photons per pulse is smaller than or equal to one, the detection probability becomes independent of previously detected events, and Eq. (6) becomes

$$\begin{aligned} P_{\text{low-flux}}(i|\boldsymbol{\lambda}) &\approx P(n>0,i|\lambda_i) \\ &= 1 - P(n=0,i|\lambda_i) \\ &= 1 - \exp(-\lambda_i), \end{aligned} \tag{8}$$

The probability $P(\mathbf{h}|\boldsymbol{\lambda})$ of detecting the full histogram $\mathbf{h}$ then simplifies to

$$\begin{aligned}
P_{\text{low-flux}}(\mathbf{h}|\boldsymbol{\lambda}) &= \prod_{i=1}^{T} \binom{N}{h_i} P_{\text{low-flux}}(i|\boldsymbol{\lambda})^{h_i} P(\text{no detections})^{N-\mathbf{1}^T\mathbf{h}} \\
&= \prod_{i=1}^{T} \binom{N}{h_i} (1-\exp(-\lambda_i))^{h_i} \exp(-\lambda_i)^{(N-h_i)} \\
&\approx \prod_{i=1}^{T} \exp(-\lambda_i N) \frac{(\lambda_i N)^{h_i}}{h_i!} = \prod_{i=1}^{T} \text{Poisson}(\lambda_i N),
\end{aligned} \tag{9}$$

where the binomial is approximated as a Poisson distribution in the low-flux regime following the Poisson limit theorem (cf. [1, 2, 4]). As the count in each histogram bin reduces to a Poisson-distributed random variable, independent of other histogram bins, low-flux photon efficient imaging solves an estimation problem to recover the latent counts $\boldsymbol{\lambda}$, which are subsequently used to estimate reflectivity $\alpha$ and depth $z$.

**Model Assumptions** The image formation model from Eq. 7 relies on the common assumption that SPAD measurements between laser pulses are independent. This assumption neglects *afterpulsing*, where an electron avalanche may be triggered by charge carriers generated in the device by previous laser pulses. In our experimental setup, the afterpulsing probability is around 1% and is therefore ignored.

## Probabilistic Reconstruction Method

In this section, we describe statistical estimation methods for recovering depth and albedo from SPAD measurements. The proposed model is independent of the illumination level and generalizes beyond the low-flux regime. First, we introduce a Bayesian maximum-a-posteriori (MAP) estimation approach with spatial priors on depth and reflectivity images. The resulting optimization problem is solved using a proximal optimization algorithm, which decomposes the challenging joint estimation problem into a sequence of separable problems, one with an objective involving the per-pixel likelihood, the others involving the prior terms. Next, we describe per-pixel likelihood of the latent intensity vector $\boldsymbol{\lambda}$ using the model introduced in Eq. (7). Following this, a direct estimation of the depth and albedo (equivalent to reflectance $\alpha$ in our notation) is described under the same model. Finally, we review the relationship of the proposed method to traditional approaches such as Coates pileup correction [3].

### Bayesian MAP Estimation

Thus far, we have only considered a single SPAD detector. In the following, we generalize our model to SPAD arrays or scenes scanned by sequentially scanning different scene points with a single SPAD. Assume that a scene is recorded at discrete points $(m, j)$ $\forall m \in \{1, \ldots, X\}, j \in$ $\{1, \ldots, Y\}$, acquiring a SPAD image consisting of $X \times Y$ histograms $\mathbf{h}^{(m,j)}$. Next, we describe how to recover a vectorized depth image $\mathbf{z} \in \mathbf{R}^{XY}$ and vectorized reflectivity image $\boldsymbol{\alpha} \in \mathbf{R}^{XY}$ from such histogram measurements. To stream-line notation in the following, we do not include the ambient term as an unknown. This unregularized offset term is incorporated into the solver method analogously to the the reflectivity image, as described in the main manuscript. Given histogram measurements, we adopt a Bayesian approach and solve a maximum-a-posteriori (MAP) estimation problem to recover depth $\mathbf{z}$ and reflectivity $\boldsymbol{\alpha}$ as

$$
\begin{aligned}
\mathbf{z}_{\text{MAP}}, \boldsymbol{\alpha}_{\text{MAP}} &= \underset{\mathbf{z},\boldsymbol{\alpha}}{\operatorname{argmax}} \; P(\mathbf{z}, \boldsymbol{\alpha} | \mathbf{h}^{(0,0)}, \ldots, \mathbf{h}^{(Z,Y)}) \\
&= \underset{\mathbf{z},\boldsymbol{\alpha}}{\operatorname{argmax}} \; P(\mathbf{h}^{(0,0)}, \ldots, \mathbf{h}^{(Z,Y)} | \mathbf{z}, \boldsymbol{\alpha}) P(\mathbf{z}, \boldsymbol{\alpha}) \\
&\approx \underset{\mathbf{z},\boldsymbol{\alpha}}{\operatorname{argmax}} \; \prod_{m,j} P(\mathbf{h}^{(m,j)} | \mathbf{z}_{(m,j)}, \boldsymbol{\alpha}_{(m,j)}) P(\mathbf{z}) P(\boldsymbol{\alpha}) \\
&= \underset{\mathbf{z},\boldsymbol{\alpha}}{\operatorname{argmin}} \; \sum_{m,j} - \log P\left(\mathbf{h}^{(m,j)} | \boldsymbol{\lambda}(\mathbf{z}_{(m,j)}, \boldsymbol{\alpha}_{(m,j)})\right) - \log P(\mathbf{z}) - \log P(\boldsymbol{\alpha}) \\
&= \underset{\mathbf{z},\boldsymbol{\alpha}}{\operatorname{argmin}} \; \sum_{m,j} \Lambda\left(\mathbf{h}^{(m,j)}, \mathbf{z}_{(m,j)}, \boldsymbol{\alpha}_{(m,j)}\right) + \Gamma_z(\mathbf{z}) + \Gamma_\alpha(\boldsymbol{\alpha}),
\end{aligned}
\tag{10}
$$

which maximizes the posterior over all measured histograms $\mathbf{h}^{(m,j)}$. The third row in the above equation makes the simplifying assumptions that the scene depth and reflectance priors are independent and that the likelihood of the histogram for each pixel is independent. The former assumption is a useful simplification for our reconstruction method, while the latter assumption is reasonable for the static scenes we capture.

Although the specific prior distributions for depth $\Gamma_z(\mathbf{z})$ and reflectivity $\Gamma_\alpha(\boldsymbol{\alpha})$ have not been defined yet, the objective in the last row of Eq. (10) exposes structure that can be leveraged in the optimization algorithm. The objective is a sum of three components. The first is a data fidelity term originating from the likelihood, which is separable in the pixels, and there are two separate regularizer terms for depth and reflectance, which result from the respective prior distributions. In the following, we will describe all of these three components in detail, starting with the prior terms.

**Priors on Depth and Reflectance** We assume Laplacian priors on the gradients of the depth and reflectivity images, which leads to Total Variation regularization terms $\Gamma_z, \Gamma_\alpha$ in Eq. (10)

$$
P(\mathbf{z}) \propto \exp\left( -\gamma_1 \|\nabla \mathbf{z}\|_1 \right), \qquad \Gamma_z(\mathbf{z}) = \gamma_1 \|\nabla \mathbf{z}\|_1, \tag{11}
$$

$$
P(\boldsymbol{\alpha}) \propto \exp\left( -\gamma_2 \|\nabla \boldsymbol{\alpha}\|_1 \right), \qquad \Gamma_\alpha(\boldsymbol{\alpha}) = \gamma_2 \|\nabla \boldsymbol{\alpha}\|_1, \tag{12}
$$

where $\gamma_1, \gamma_2$ are prior weights that are estimated empirically from a representative dataset of depth and albedo images, respectively.

**Likelihood** The log-likelihood term in Eq. (10) is separable in the measurements $(m, j)$. With the arrival rate vector $\boldsymbol{\lambda}$ from Eq. (2) a function of $\alpha$, $z$, we can expand $\Lambda(\mathbf{h}, z, \alpha)$ as follows

$$\begin{aligned}
\Lambda(\mathbf{h}, z, \alpha) &= -\log P\left(\mathbf{h}|\boldsymbol{\lambda}(z,a)\right) \qquad \text{with} \qquad \lambda_i(z,a) = \mu\left(\alpha\, \tilde{g}(t_i - 2z/c) + s\right) + d \\
&= \mathbf{1}^T \boldsymbol{\lambda}(z,a)(N - \mathbf{1}^T\mathbf{h}) - \sum_{i=1}^{T} \mathbf{h}_i \log\left(e^{-\sum_{k=1}^{i-1} \lambda_k(z,a)} - e^{-\sum_{k=1}^{i} \lambda_k(z,a)}\right) \\
&= \mathbf{1}^T \boldsymbol{\lambda}(z,a)(N - \mathbf{1}^T\mathbf{h}) + \sum_{i=1}^{T} \mathbf{h}_i \sum_{k=1}^{i-1} \lambda_k(z,a) - \sum_{i=1}^{T} \mathbf{h}_i \log\left(1 - e^{-\lambda_i(z,a)}\right),
\end{aligned} \tag{13}$$

where we use the logarithmic identity $\log(a - b) = \log(a) + \log(1 - b/a)$ in the last row.

### Proximal Optimization

Having defined the objective function from Eq. (10), we now describe the algorithm to solve this minimization problem. We propose an efficient proximal optimization method that leverages the separable structure of the objective from Eq. (10) in the measurements $(m, j)$. Following [5, 6] we reformulate the objective by splitting the unknowns in the three components while enforcing consensus in the constraints

$$\begin{aligned}
& \text{minimize} \; \sum_{m,j} \Lambda\left(\mathbf{h}^{(m,j)}, \mathbf{z}_{(m,j)}, \boldsymbol{\alpha}_{(m,j)}\right) \; + \; \gamma_1 \|\nabla \mathbf{z}\|_1 \; + \; \gamma_2 \|\nabla \boldsymbol{\alpha}\|_1 \\
= \; & \text{minimize} \; \underbrace{\sum_{m,j} \Lambda\left(\mathbf{h}^{(m,j)}, \mathbf{x}_{(m,j)}\right)}_{g(\mathbf{x})} \; + \; \underbrace{\gamma_1 \|\mathbf{y}_1\|_1}_{f_1(\mathbf{y}_1)} \; + \; \underbrace{\gamma_2 \|\mathbf{y}_2\|_1}_{f_2(\mathbf{y}_2)} \\
& \text{subject to} \quad \underbrace{\begin{bmatrix} \nabla & \\ & \nabla \end{bmatrix}}_{\mathbf{K}} \underbrace{\begin{bmatrix} \mathbf{z} \\ \boldsymbol{\alpha} \end{bmatrix}}_{\mathbf{x}} = \underbrace{\begin{bmatrix} \mathbf{y}_1 \\ \mathbf{y}_2 \end{bmatrix}}_{\mathbf{y}} \\
= \; & \text{minimize} \; g(\mathbf{x}) \; + \; f_1(\mathbf{y}_1) \; + \; f_2(\mathbf{y}_2) \\
& \text{subject to} \quad \mathbf{K}\mathbf{x} = \mathbf{y},
\end{aligned}$$

where we stack the unknowns $\mathbf{z}, \boldsymbol{\alpha}$ in the vector $\mathbf{x} = [\mathbf{z}^T, \boldsymbol{\alpha}^T]^T$. We have introduced two slack variables $\mathbf{y}_{1,2} \in \mathbf{R}^{2XY}$ for the gradients of the depth and albedo images $\nabla \mathbf{z}, \nabla \boldsymbol{\alpha} \in \mathbf{R}^{2XY}$, and enforce consensus between the gradients and the slack variables in the constraint. Furthermore, we have introduced functions $g$ and $f_{1,2}$ for the individual objective components in Eq. (14).

We solve this objective using a linearized variant of the Alternating Method of Multipliers (ADMM) [5, 6]. To this end, we first formulate the augmented Lagrangian of Eq. (14) as

$$L_\rho(\mathbf{x}, \mathbf{y}, \mathbf{u}) = g(\mathbf{x}) + \sum_{i=1}^{2} f_i(\mathbf{y}_i) + \frac{\rho}{2} \|\mathbf{K}\mathbf{x} - \mathbf{y} + \mathbf{u}\|_2^2, \tag{14}$$

which introduces here a scaled Lagrange variable $\mathbf{u}$. The ADMM algorithm solves the saddle-point problem of minimizing $L_\rho$ w.r.t. each primal variable $\mathbf{x}, \mathbf{y}$, while maximizing the $L_\rho$ w.r.t. $\mathbf{u}$. In block-coordinate descent fashion, a single ADMM iteration minimizes $\mathbf{x}$ and $y$ in two separate sub-steps, and performs a dual ascent with the step size $\rho$ as the last step.

Linearized ADMM is a variant of the standard ADMM algorithm, where we replace the term

$$(\rho/2)\|\mathbf{K}\mathbf{x} - \mathbf{y}^k + \mathbf{u}^k\|_2^2 \tag{15}$$

in the augmented Lagrangian Eq. (14) with its linearization plus quadratic regularization:

$$\rho\mathbf{K}^T(\mathbf{K}\mathbf{x} - \mathbf{y}^k + \mathbf{u}^k) + (\mu/2)\|\mathbf{x} - \mathbf{x}^k\|_2^2. \tag{16}$$

Performing dual ascent on the linearized augmented Lagrangian results in Algorithm 1.

**Algorithm 1** Linearized ADMM to solve Problem (14)

Initialization: $\mu > \rho\|\mathbf{K}\|_2^2$, $(\mathbf{x}^0, \mathbf{y}^0, \mathbf{u}^0)$.
**for** $k = 1$ to $V$ **do**
  $\mathbf{x}^{k+1} = \mathbf{prox}_{\frac{g}{\mu}}(\mathbf{x}^k - (\rho/\mu)\mathbf{K}^T(\mathbf{K}\mathbf{x}^k - \mathbf{y}^k + \mathbf{u}^k))$ Likelihood Update (17)
  $\mathbf{y}_j^{k+1} = \mathbf{prox}_{\frac{f_j}{\rho}}(\mathbf{K}_j\mathbf{x}_j^{k+1} + \mathbf{u}_j^k) \quad \forall j \in \{1, 2\}$ Prior Updates (18)
  $\mathbf{u}_j^{k+1} = \mathbf{u}_j^k + (\mathbf{K}_j\mathbf{x}^{k+1} - \mathbf{y}_j^{k+1}) \quad \forall j \in \{1, 2\}$
**end for**

This algorithm contains two minimization steps, one w.r.t $\mathbf{x}$ in step (17), and another for $y$ in step (18). Both minimization steps are expressed using proximal operators as a common notational shorthand [5], which is defined as

$$\mathbf{prox}_{\phi f}(\mathbf{v}) = \underset{\mathbf{x}}{\operatorname{argmin}} \left( f(\mathbf{x}) + \frac{1}{2\phi}\|\mathbf{x} - \mathbf{v}\|_2^2 \right),$$

where $\phi > 0$ and $\mathbf{v} \in \mathbf{R}^{XY}$. Proximal operators are alternatives to classical gradient and Hessian oracles used by first and second order optimization methods. Algorithm 1 applies separate proximal operators for each $f_{1,2}$ as the objective Eq. (14) is separable in the two components of $\mathbf{y}$. Note that for convex objectives the algorithm is identical to the Pock-Chambolle algorithm [7], which is in fact linearized ADMM applied to the dual problem. For nonconvex objectives, such as Eq. 13, the two algorithms may give different results. Next, we specify and describe all relevant algorithm parameter choices and termination criteria.

**Algorithm Parameters and Termination Criteria** For Algorithm 1 we use the parameter choices $\rho = 1/\|\mathbf{K}\|_2$, $\mu = \|\mathbf{K}\|_2$, $\mathbf{x}^0 = 0$, $\mathbf{y}^0 = 0$, and $\mathbf{u}^0 = 0$. We use the termination the criteria from [6], which stops the algorithm when the norms of the primal residual $r^{k+1} = \mathbf{K}\mathbf{x}^{k+1} - \mathbf{y}^{k+1}$ and dual residual $s^{k+1} = \rho\mathbf{K}^T(\mathbf{y}^{k+1} - \mathbf{y}^k)$ fall below given thresholds. The threshold for the primal residual is

$$\epsilon^{\text{pri}} = \sqrt{m}\epsilon^{\text{abs}} + \epsilon^{\text{rel}} \max\{\|\mathbf{K}\mathbf{x}^k\|_2, \|\mathbf{y}^k\|_2\},$$

where we use $\epsilon^{\text{abs}} = \epsilon^{\text{rel}} = 10^{-4}$. The dual threshold is

$$\epsilon^{\text{pri}} = \sqrt{n}\epsilon^{\text{abs}} + \epsilon^{\text{rel}}\|\mathbf{K}^T\mathbf{u}^k\|_2.$$

**Proximal Prior Update in Eq. (18)** Evaluating the two proximal operator updates from step (18) reduces to the well-known shrinkage operator [5], that is

$$\mathbf{prox}_{\frac{f_j}{\nu}}(\mathbf{v}) = \mathbf{prox}_{\frac{\gamma_j}{\nu}\|\cdot\|_1}(\mathbf{v}) = \max\left(1 - \frac{\gamma_j}{\nu|\mathbf{v}|}, 0\right) \odot \mathbf{v}, \qquad \text{Shrinkage (19)}$$

which is a point-wise function that can be efficiently evaluated in parallel for each vector component (i.e., each pixel $(m, j)$).

**Likelihood Proximal Update in Eq. (17)** The proximal operator update in Eq. 17 is given by

$$\begin{aligned} \mathbf{prox}_{\frac{g}{\mu}}(\mathbf{v}) &= \underset{\mathbf{x}}{\operatorname{argmin}}\ g(\mathbf{x}) + \frac{\mu}{2}\|\mathbf{x} - \mathbf{v}\|_2^2 \\ &= \underset{\mathbf{x}}{\operatorname{argmin}}\ \sum_{m,j} \Lambda\big(\mathbf{h}^{(m,j)}, \mathbf{x}_{(m,j)}\big) + \frac{\mu}{2}\|\mathbf{x} - \mathbf{v}\|_2^2 \qquad (20) \\ \Leftrightarrow\ \Big(\mathbf{prox}_{\frac{g}{\mu}}(\mathbf{v})\Big)_{(m,j)} &= \underset{z,\alpha}{\operatorname{argmin}}\ \Lambda\big(\mathbf{h}^{(m,j)}, z, \alpha\big) + \frac{\mu}{2}\left(z - \mathbf{v}^z_{(m,j)}\right)^2 + \frac{\mu}{2}\left(\alpha - \mathbf{v}^\alpha_{(m,j)}\right)^2 \quad \forall (m,j), \end{aligned}$$

where $\mathbf{v}^z_{(m,j)}, \mathbf{v}^\alpha_{(m,j)} \in \mathbf{R}$ extract the depth and reflectivity of pixel $(m, j)$ and $z, \alpha \in \mathbf{R}$ are the optimization variables. The form of the update reveals why we linearized the augmented Lagrangian in the proposed optimization method: the minimization for depth and reflectance is separable in the pixels $(m, j)$. Unlike Eq. 18, the proximal update cannot be expressed in closed form. Instead we compute the joint optimization over all histograms as a sequence of bi-variate non-linear optimization problems that can be solved in parallel for all pixels $(m, j)$. In the following, we describe how each of these non-linear problems is solved.

**Bivariate Per-Pixel Likelihood Optimization** Given pixel $(m, j)$, we denote the corresponding histogram $\mathbf{h}$ and proximal slack terms $\mathbf{v}^z, \mathbf{v}^\alpha$ for notational simplicity below. Recalling Eq. (13), we solve the following minimization problem jointly for $z, \alpha$:

$$\text{minimize}\ \underbrace{\Lambda(\mathbf{h}, z, \alpha) + \frac{\mu}{2}(z - \mathbf{v}^z)^2 + \frac{\mu}{2}(\alpha - \mathbf{v}^\alpha)^2}_{\Upsilon(z,\alpha)} \qquad (21)$$

$$\begin{aligned} = \text{minimize}\ &\underbrace{\mathbf{1}^T\boldsymbol{\lambda}(z,\alpha)(N - \mathbf{1}^T\mathbf{h}) + \sum_{i=1}^{T}\mathbf{h}_i\sum_{k=1}^{i-1}\lambda_k(z,\alpha) - \sum_{i=1}^{T}\mathbf{h}_i\log\left(1 - e^{-\lambda_i(z,\alpha)}\right)}_{\tilde{\Lambda}(\boldsymbol{\lambda}(z,\alpha))}, \qquad (22) \\ &+ \frac{\mu}{2}(z - \mathbf{v}_z)^2 + \frac{\mu}{2}(\alpha - \mathbf{v}_\alpha)^2 \end{aligned}$$

which introduces a bivariate function $\Upsilon$ that is the sum of the composite function $\tilde{\Lambda}\left(\boldsymbol{\lambda}(z,\alpha)\right)$ and quadratic proximal closeness terms. We solve the non-linear bivariate minimization problem with the Newton Method given in Algorithm 2. We terminate this iteration once

**Algorithm 2** Bivariate Newton Method to solve Problem (21)

1: Initialization: $z^0 = \mathbf{v}_z$, $\alpha^0 = \mathbf{v}_\alpha$.
2: **for** $k = 1$ to $V$ **do**
3: $\begin{bmatrix} z^{k+1} \\ \alpha^{k+1} \end{bmatrix} = \begin{bmatrix} z^{k} \\ \alpha^{k} \end{bmatrix} - \left[\mathbf{H}\Upsilon(z^k,\alpha^k)\right]^{-1} \begin{bmatrix} \frac{\partial \Upsilon(z^k,\alpha^k)}{\partial z} \\ \frac{\partial \Upsilon(z^k,\alpha^k)}{\partial \alpha} \end{bmatrix}$ Newton Update (23)
4: **end for**

optimality is achieved and the gradient norm $\left\|[\partial\Upsilon/\partial z, \partial\Upsilon/\partial\alpha]^T\right\|_2 < \epsilon_{\text{newton}}$ falls below the threshold $\epsilon_{\text{newton}} = 10^{-4}$. The Newton update step (23) requires evaluation of the gradient $\nabla\Upsilon$ and Hessian $H\Upsilon$. Both are derived in the following.

**Gradient of Proximal Likelihood in Eq. (23)** We use the chain rule to derive the gradient of the composite part of $\Upsilon$ as

$$\begin{aligned} \begin{bmatrix} \frac{\partial \Upsilon(z,\alpha)}{\partial z} \\ \frac{\partial \Upsilon(z,\alpha)}{\partial \alpha} \end{bmatrix} &= \nabla\left(\tilde{\Lambda}\circ\boldsymbol{\lambda}\right)(z,\alpha) + \mu \begin{bmatrix} z - \mathbf{v}_z \\ \alpha - \mathbf{v}_\alpha \end{bmatrix} \\ &= \left(J\boldsymbol{\lambda}\left(z,\alpha\right)\right)^T \nabla\tilde{\Lambda}(\boldsymbol{\lambda}\left(z,\alpha\right)) + \mu \begin{bmatrix} z - \mathbf{v}_z \\ \alpha - \mathbf{v}_\alpha \end{bmatrix} \\ &= \left[\frac{\partial\boldsymbol{\lambda}(z,\alpha)}{\partial z}\ \frac{\partial\boldsymbol{\lambda}(z,\alpha)}{\partial \alpha}\right]^T \nabla\tilde{\Lambda}(\boldsymbol{\lambda}\left(z,\alpha\right)) + \mu \begin{bmatrix} z - \mathbf{v}_z \\ \alpha - \mathbf{v}_\alpha \end{bmatrix}, \end{aligned} \tag{24}$$

where $J$ denotes here the Jacobian matrix. The gradient of $\tilde{\Lambda}$ can be derived directly from Eq. (22) and defined in terms of its partials as

$$\frac{\partial\tilde{\Lambda}}{\partial\boldsymbol{\lambda}_i} = (N - \sum_{k=1}^{i}\mathbf{h}_k) - \frac{\mathbf{h}_i e^{-\boldsymbol{\lambda}_i}}{1 - e^{-\boldsymbol{\lambda}_i}} \qquad \forall i \in \{1,\ldots,T\}. \tag{25}$$

The Jacobian of the arrival rates can be derived using Eqs. (2) and (3), defined in terms of its partials as

$$\frac{\partial\boldsymbol{\lambda}_i}{\partial z} = \frac{2\mu\alpha}{c}\sum_{k=1}^{K} a_k e^{-\frac{(t_i - 2z/c - b_k)^2}{c_k^2}} \cdot \left(2\frac{(t_i - 2z/c - b_k)}{c_k^2}\right) \qquad \forall i \in \{1,\ldots,T\}, \tag{26}$$

$$\frac{\partial\boldsymbol{\lambda}_i}{\partial \alpha} = \mu\sum_{k=1}^{K} a_k e^{-\frac{(t_i - 2z/c - b_k)^2}{c_k^2}} \qquad \forall i \in \{1,\ldots,T\}. \tag{27}$$

**Hessian of Proximal Likelihood in Eq. (23)** In similar fashion to the gradient derivations above, we apply the chain rule to estimate the $2 \times 2$ Hessian matrices for each pixel.

Specifically, it is

$$\mathbf{H}\Upsilon(z,\alpha) = \left(J\boldsymbol{\lambda}\left(z,\alpha\right)\right)^T \, \mathbf{H}\tilde{\Lambda} \, \left(J\boldsymbol{\lambda}\left(z,\alpha\right)\right) \; + \; (\nabla\tilde{\Lambda})^T \otimes \mathbf{H}\boldsymbol{\lambda}(z,\alpha), \tag{28}$$

where the Jacobian matrices have already been defined in Eq. (26). The Hessian matrix $\mathbf{H}\tilde{\Lambda}$ is a diagonal matrix with

$$\frac{\partial^2 \tilde{\Lambda}}{\partial^2 \boldsymbol{\lambda}_i} = \frac{\mathbf{h}_i e^{-\boldsymbol{\lambda}_i}}{(1-e^{-\boldsymbol{\lambda}_i})^2} \qquad \forall i \in \{1,\ldots,T\}, \tag{29}$$

$$\frac{\partial^2 \tilde{\Lambda}}{\boldsymbol{\lambda}_i \boldsymbol{\lambda}_j} = 0 \qquad \forall i \neq j. \tag{30}$$

Finally, the Hessian $\mathbf{H}\boldsymbol{\lambda}(z,\alpha)$ can be derived immediately from Eqs.(2) and (3) as

$$\mathbf{H}\boldsymbol{\lambda}(z,\alpha) = \begin{bmatrix} \frac{\partial^2 \boldsymbol{\lambda}(z,\alpha)}{\partial^2 z} & \frac{\partial^2 \boldsymbol{\lambda}(z,\alpha)}{\partial z \partial \alpha} \\ \frac{\partial^2 \boldsymbol{\lambda}(z,\alpha)}{\partial z \partial \alpha} & \frac{\partial^2 \boldsymbol{\lambda}(z,\alpha)}{\partial^2 \alpha} \end{bmatrix} \quad \text{with}$$

$$\begin{aligned}
\frac{\partial^2 \boldsymbol{\lambda}_i(z,\alpha)}{\partial^2 z} &= \frac{4\mu\alpha^2}{c^2} \sum_{k=1}^{K} a_k e^{-\frac{(t_i - 2z/c - b_k)^2}{c_k^2}} \cdot \left( \frac{4(t_i - 2z/c - b_k)^2}{c_k^4} - \frac{2}{c_k^2} \right) && \forall i \in \{1,\ldots,T\}, \\
\frac{\partial \boldsymbol{\lambda}_i(z,\alpha)}{\partial \alpha} &= \frac{2\mu}{c} \sum_{k=1}^{K} a_k e^{-\frac{(t_i - 2z/c - b_k)^2}{c_k^2}} \cdot \left( \frac{2(t_i - 2z/c - b_k)}{c_k^2} \right) && \forall i \in \{1,\ldots,T\}, \\
\frac{\partial^2 \boldsymbol{\lambda}_i(z,\alpha)}{\partial^2 \alpha} &= 0.
\end{aligned} \tag{31}$$

By reusing the partials and function evaluations for the gradients, the Hessian can be computed efficiently along with the gradient evaluations in the Newton update step. We run Newton iterations in parallel for each pixel, which can be implemented very efficiently using modern graphics processing units (GPUs).

## Empirical Verification

We validate the proposed probabilistic image formation model and reconstruction method by acquiring raw histogram measurements at different incident intensities. We use a scene consisting of a single diffuse reflector at a distance of 1 m from the detector and laser. To control the incident intensity, and thus the magnitude of the pileup effect, we place neutral-density (ND) filters of various optical densities in front of the detector. Using a low-transmission (1%) ND filter allows us to enforce a low-flux regime where pileup can be eliminated, though at the cost of a 100× increased acquisition time. Even when the measurements are distorted by pileup, the proposed reconstruction method is able to fit a model which closely matches a low-flux reference measurement (see Fig. 1). The individual plots in the figure validate that the proposed image formation model and the probabilistic reconstruction generalize across different impulse response functions. In all cases, the proposed method estimates a model that matches the *unseen* reference measurement with high accuracy.

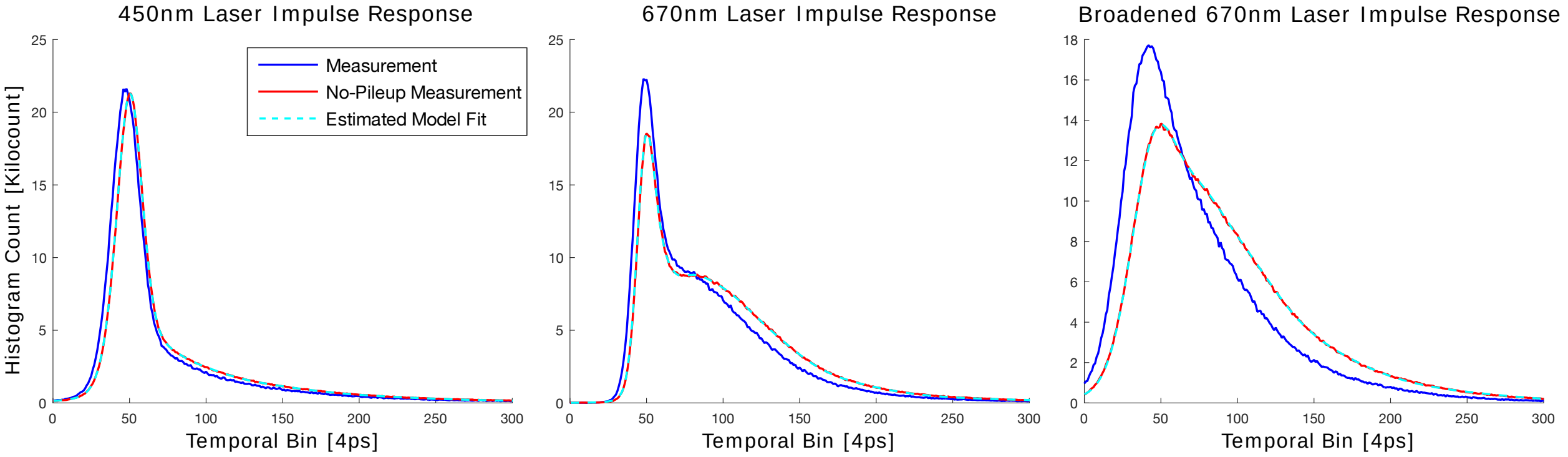

**Supplementary Figure 1:** Verification of the proposed forward model and reconstruction method. The three plots show measurements for the 670 nm Alphalas LD-670-50 laser (FWHM of 50 ps), 450 nm Alphalas LD-450-50 laser (FWHM of 90 ps), and pulse-broadened 450 nm Alphalas LD-450-50 laser (broadened to 200 ps). Each plot displays raw histogram counts in blue, acquired without an ND filter, and a corresponding low-flux measurement captured with a 1% ND filter with negligible pileup. We show the model $\tilde{g}$ with coefficients recovered from the distorted raw counts using the proposed method. For all impulse response functions, the estimated model fits the unseen reference measurement with low residual error.

### Relation to Coates Method

Coates method is a classical approach to pile-up correction [3]. It is a special case of our framework where we solve for $\boldsymbol{\lambda}$ rather than $\mathbf{z}$ and $\boldsymbol{\alpha}$ with the prior

$$P(\boldsymbol{\lambda}) = \begin{cases} 1, & \text{if } \boldsymbol{\lambda}_i \geq 0, \text{for all } i \\ 0, & \text{otherwise.} \end{cases}$$

The log-likelihood in Eq.10 is convex in $\boldsymbol{\lambda}$ with the above prior, so we find the optimum by setting the gradient in Eq. 25 to zero, yielding

$$\boldsymbol{\lambda}_i = -\log\left(1 - \frac{\mathbf{h}_i}{N - \sum_{k=1}^{i-1} \mathbf{h}_k}\right). \tag{32}$$

After applying Coates method to recover $\lambda$, we can estimate depth $\mathbf{z}$ and reflectivity $\boldsymbol{\alpha}$ using a variety of methods, including those used in the low-flux regime [2, 4].

## Additional Calibration Details

This section provides additional detail on the calibration procedure used to estimate parameters in the image formation model, described in the corresponding section. We assume that the detector's photon detection probability $\mu$ and dark count rate $d$ have been calibrated independently of the setup discussed in this work, using established calibration methods for single photon detectors (see, for example, Polyakov [8]). The remaining unknowns of the proposed model are the coefficients $\{a_k, b_k, c_k | k \in 1, \ldots, K\}$ of the Gaussian mixture model for the overall system impulse response $\tilde{g}$ from Eq. (3).

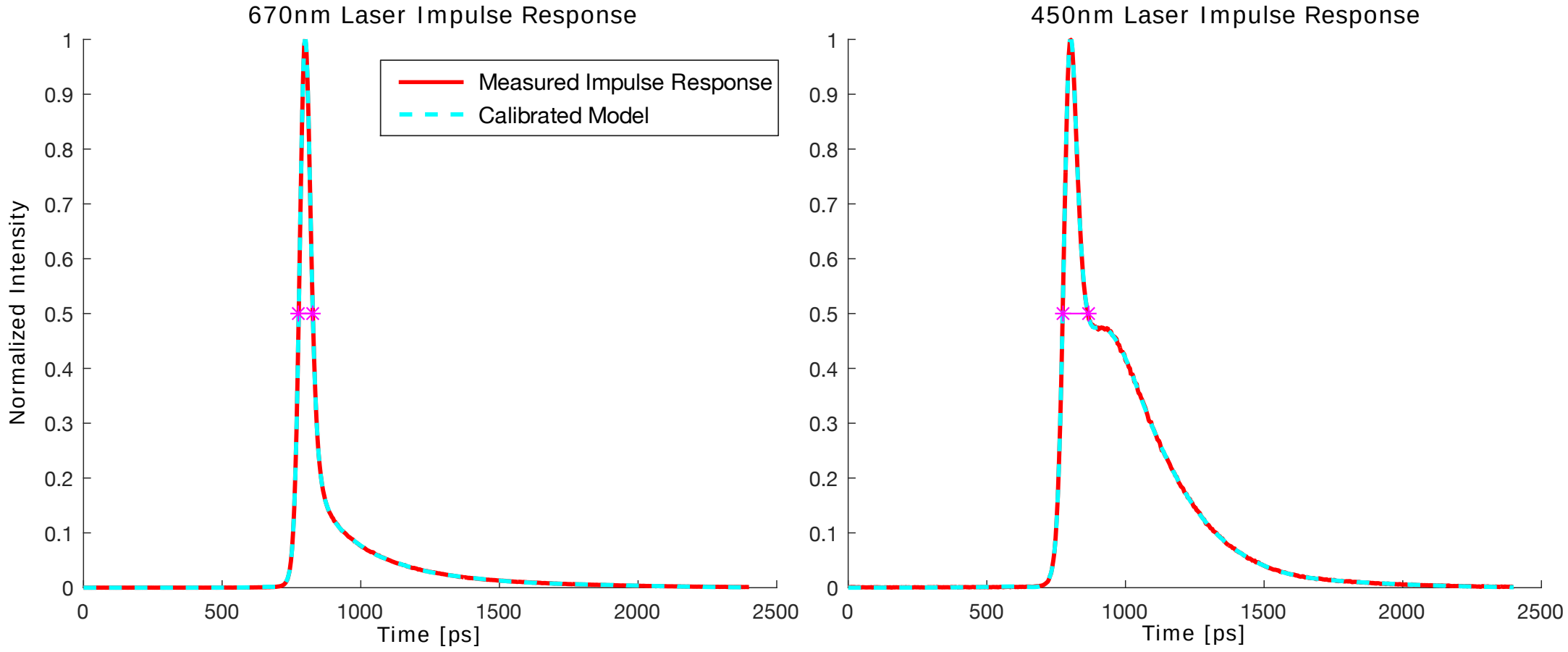

**Supplementary Figure 2:** Calibration of the laser impulse response function. The left plot shows the impulse response of the red Alphalas LD-670-50 laser (FWHM of 50 ps in magenta) and the right plot shows the impulse response of the blue Alphalas LD-450-50 laser (FWHM of 90 ps). The blue plot shows the respective raw histogram counts captured with a 1% transmission ND filter to eliminate pileup. The respective FWHM is indicated by a magenta line. The red line shows the fit of the model $\tilde{g}(\tau)$ from the Image Formation Model section with $K = 8$ Gaussian components which exhibits only low residual error.

To estimate these parameters, we acquire measurements of a scene consisting of a single diffuse reflector placed at 1 m distance, similar to the validation of the proposed forward model described at the end of the Image Formation Model section. We place an ND filter with 1% transmission in front of the laser illumination and capture a low-noise histogram of a single reflector point by accumulating $N = 10^9$ pulses, which requires approximately 17 minutes at a 1 MHz laser repetition rate.

The high-absorption ND filter ensures a low-flux regime where pileup can be ignored and where the histogram counts are Poisson distributed, as described in the Image Formation Model section. Moreover, for the extremely high numbers of shots, the Poisson distributed histogram counts become Gaussian distributed (the limiting form of the Poisson distribution is the Gaussian distribution). This allows us to rely on efficient conventional expectation-maximization algorithms for normal mixture models in the proposed calibration procedure. For further details on this algorithms, we refer the interested reader to Murphy [9]. Finally, Figure 2 shows calibration results for two different short-pulsed lasers used in this work. In both cases, fitting $K = 8$ mixtures allows for a high-fidelity impulse response model.

# Supplementary Results

## Evaluation in Simulation

In the following, we evaluate the proposed method using simulated measurements with known ground-truth depth and reflectivity. Moreover, in addition to quantifying reconstruction error on a wide variety of scenes, synthesizing observations allows us also to analyze the effect of acquisition parameters on the accuracy of the proposed and of previously proposed reconstruction methods.

### Depth and Reflectivity Error Evaluation

To evaluate the performance of the proposed method in representative real-world scenarios, we use ground-truth depth and intensity images from the Middlebury Stereo dataset [10] which includes scenes of highly varying shape and relative distances. However, the intensity images in this data set are captured under the respective scene illumination, not under active illumination comparable to the proposed setup. To account for this, we estimate scene albedo value for the scene using the approach from Chen and Koltun [11].

Given a ground-truth depth/reflectance map pair, we simulate histogram measurements using the full probabilistic image formation model for every pixel with $N = 10,000$ shots. In the following, we show depth and albedo reconstruction error results for three different laser impulse response functions: the calibrated blue Alphalas LD-450-50 and red Alphalas LD-670-50 lasers from Fig. 2, and a synthetic, idealized Gaussian impulse response (FWHM of 50 ps). Figs. 3-8 show the depth and albedo error per scene for all impulse response functions. Here, we compare the proposed method against the conventional log-matched filter estimate [2], Gaussian fit [12], first-photon imaging [4, 1], and Coates method [3] followed by a Gaussian fit.

The logarithmic scale in the depth errors reveals that the proposed method outperforms all competing methods by two orders of magnitude in depth accuracy. Similarly, the reflectance estimates show an order of magnitude in improvement over all existing methods. Hence, the proposed method opens the space of sub-picosecond accurate imaging which has not been possible with previous approaches. Tables 1, 2, and 3 show the average depth error and reflectance PSNR per scene for all three laser impulse response functions, the calibrated blue and red laser responses from Fig. 2, and the synthetic Gaussian impulse response. For compactness, we only include the conventional log-matched filter method, Coates [3] pileup correction followed by a Gaussian fit, and the first-photon imaging method [4, 1]. All error tables consistently verify that, independent of the scene and laser impulse response, we achieve two orders of magnitude improved depth accuracy and an order of magnitude lower reflectance error compared with the best existing method.

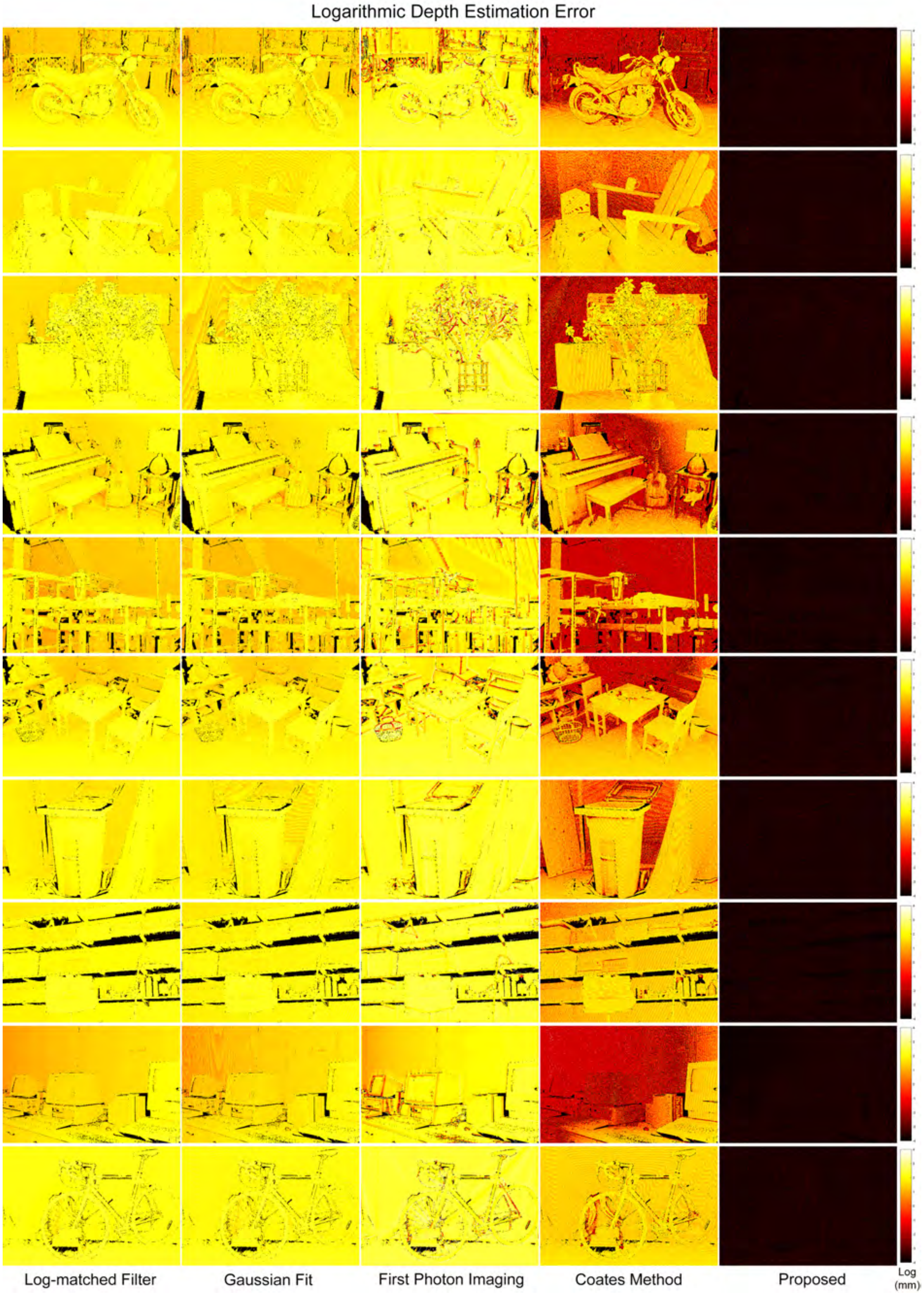

**Supplementary Figure 3:** Depth Error Maps for the Alphalas LD-670-50 laser (FWHM of 50 ps). We compare the conventional log-matched filter estimate [2], Gaussian fit [12], first-photon imaging [4, 1], and Coates [3] followed by Gaussian fit.

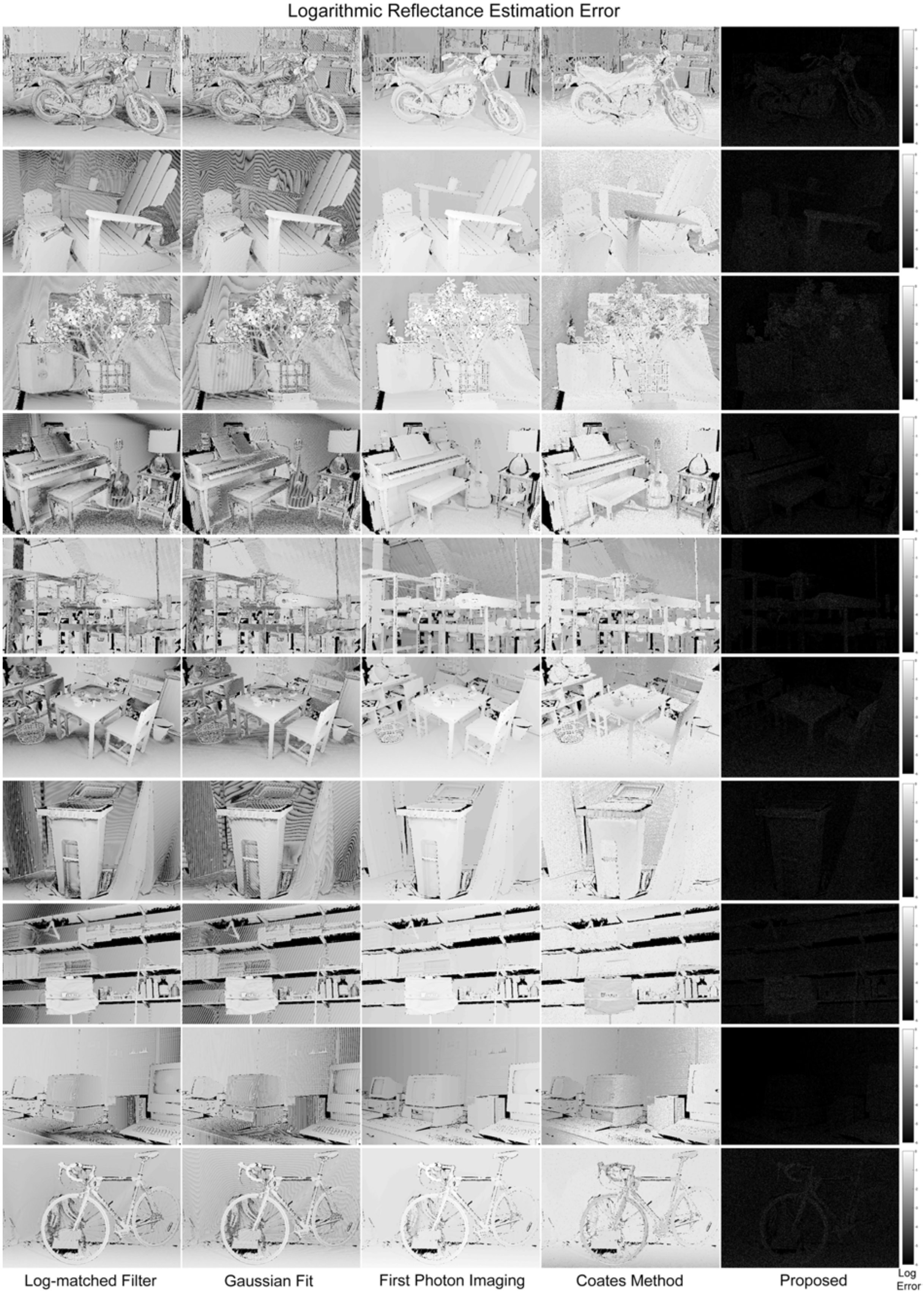

**Supplementary Figure 4:** Reflectance Error Maps for the Alphalas LD-670-50 laser. Scenes from the Middlebury Stereo dataset [10] are shown.

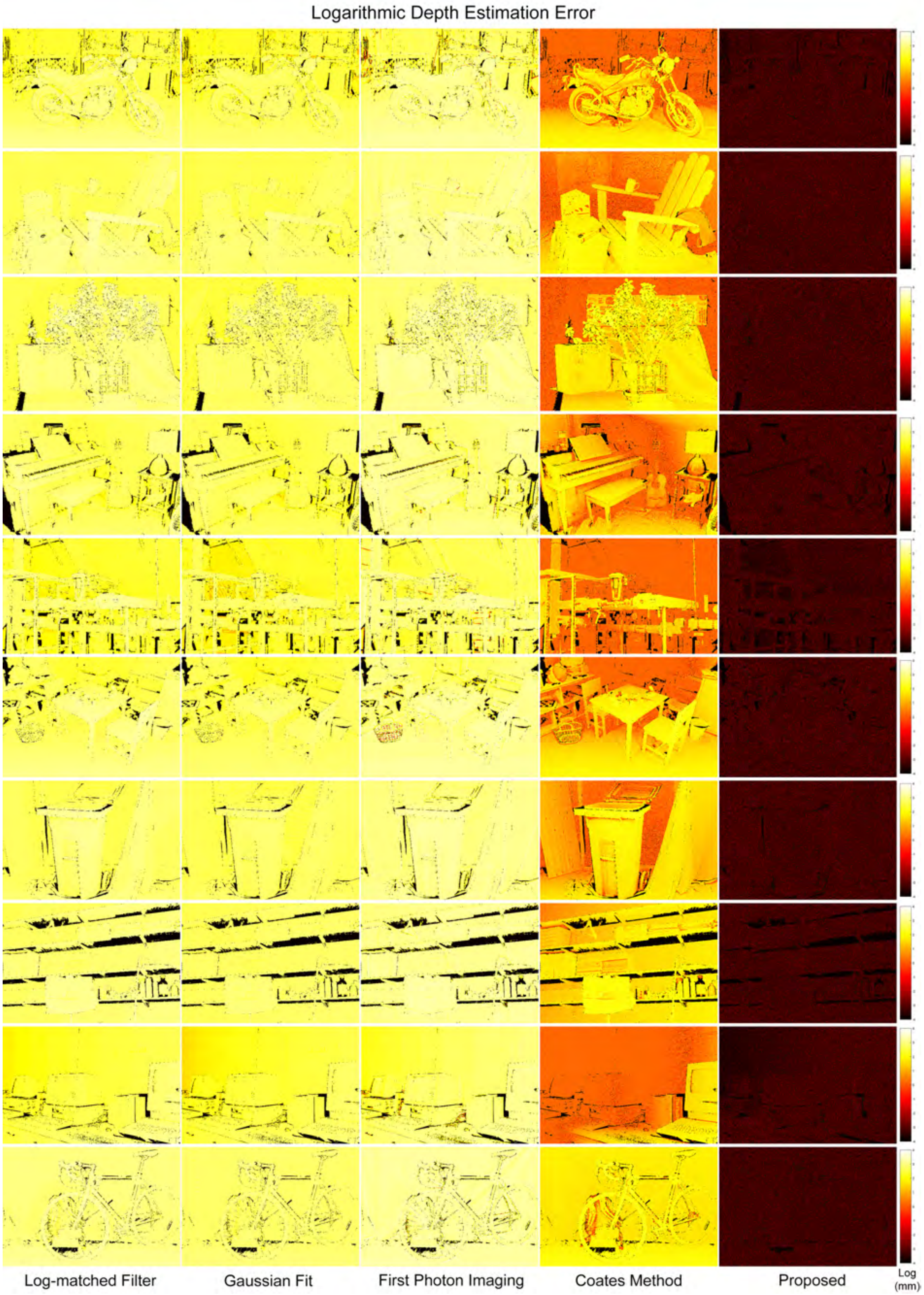

**Supplementary Figure 5:** Depth Error Maps for the Alphalas LD-450-50 laser (FWHM of 90 ps). Scenes from the Middlebury Stereo dataset [10] are shown.

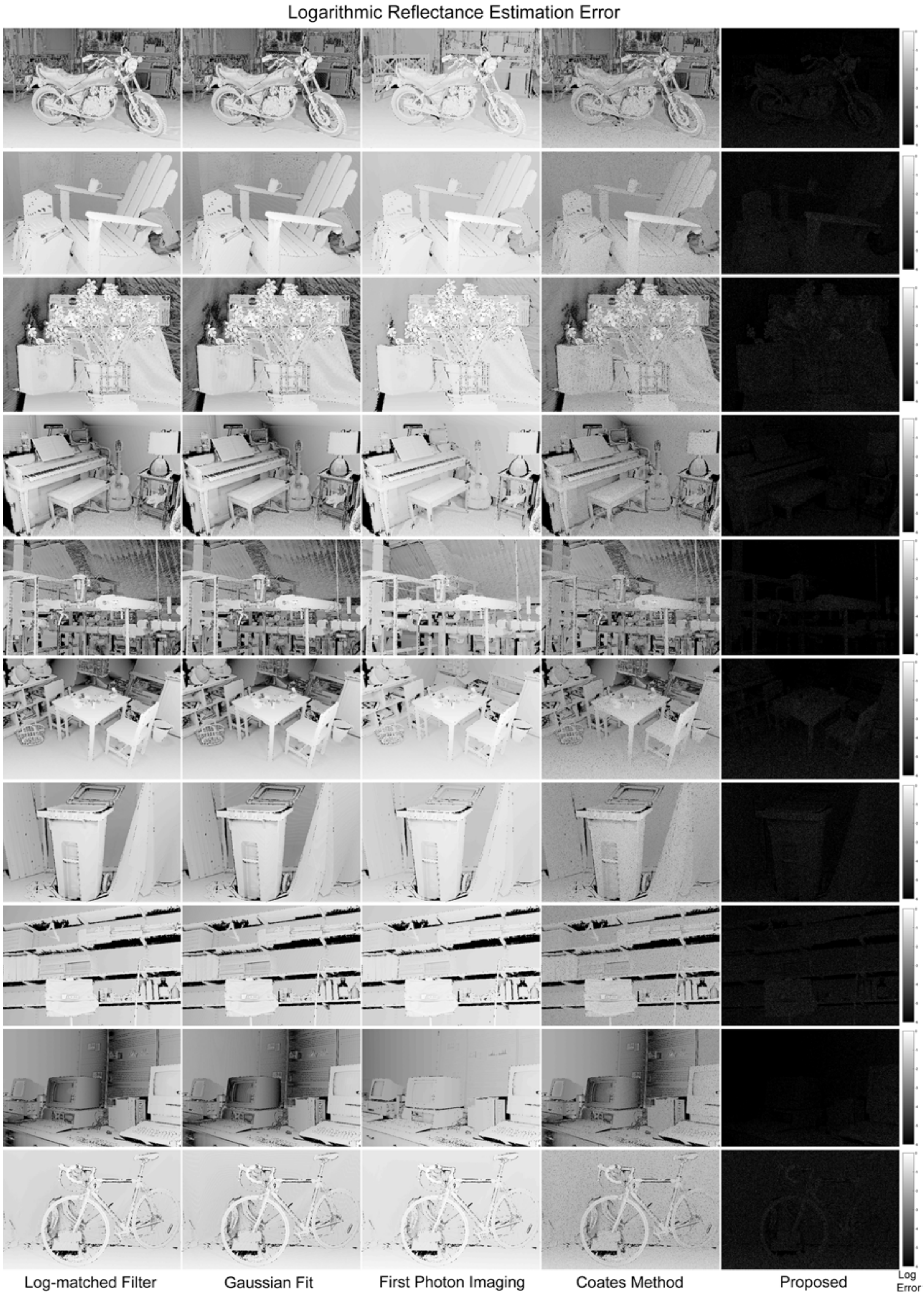

**Supplementary Figure 6:** Reflectance Error Maps for the Alphalas LD-450-50 laser. Scenes from the Middlebury Stereo dataset [10] are shown.

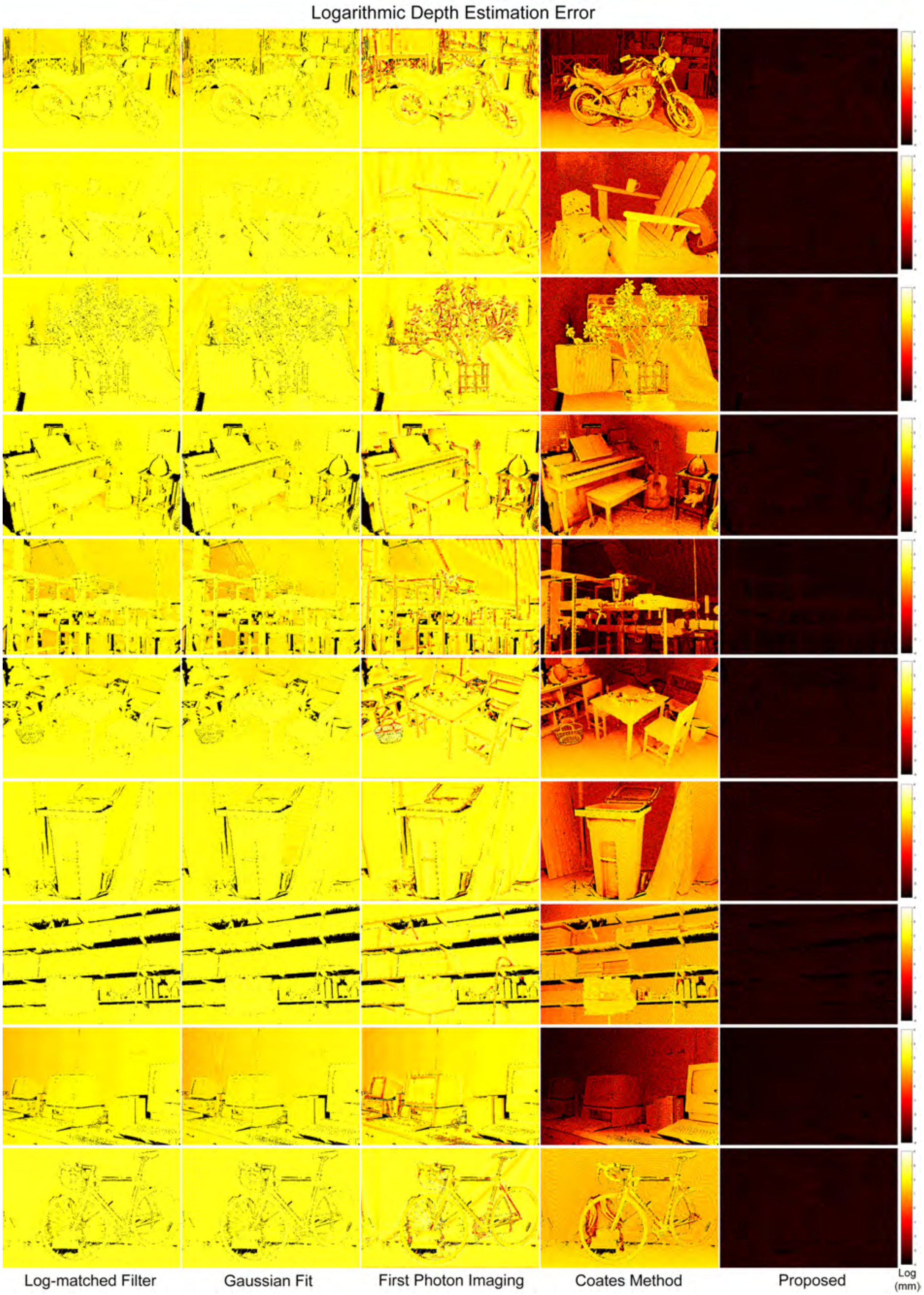

**Supplementary Figure 7:** Depth Error Maps for an ideal Gaussian laser impulse response (FWHM of 50 ps). Scenes from the Middlebury Stereo dataset [10].

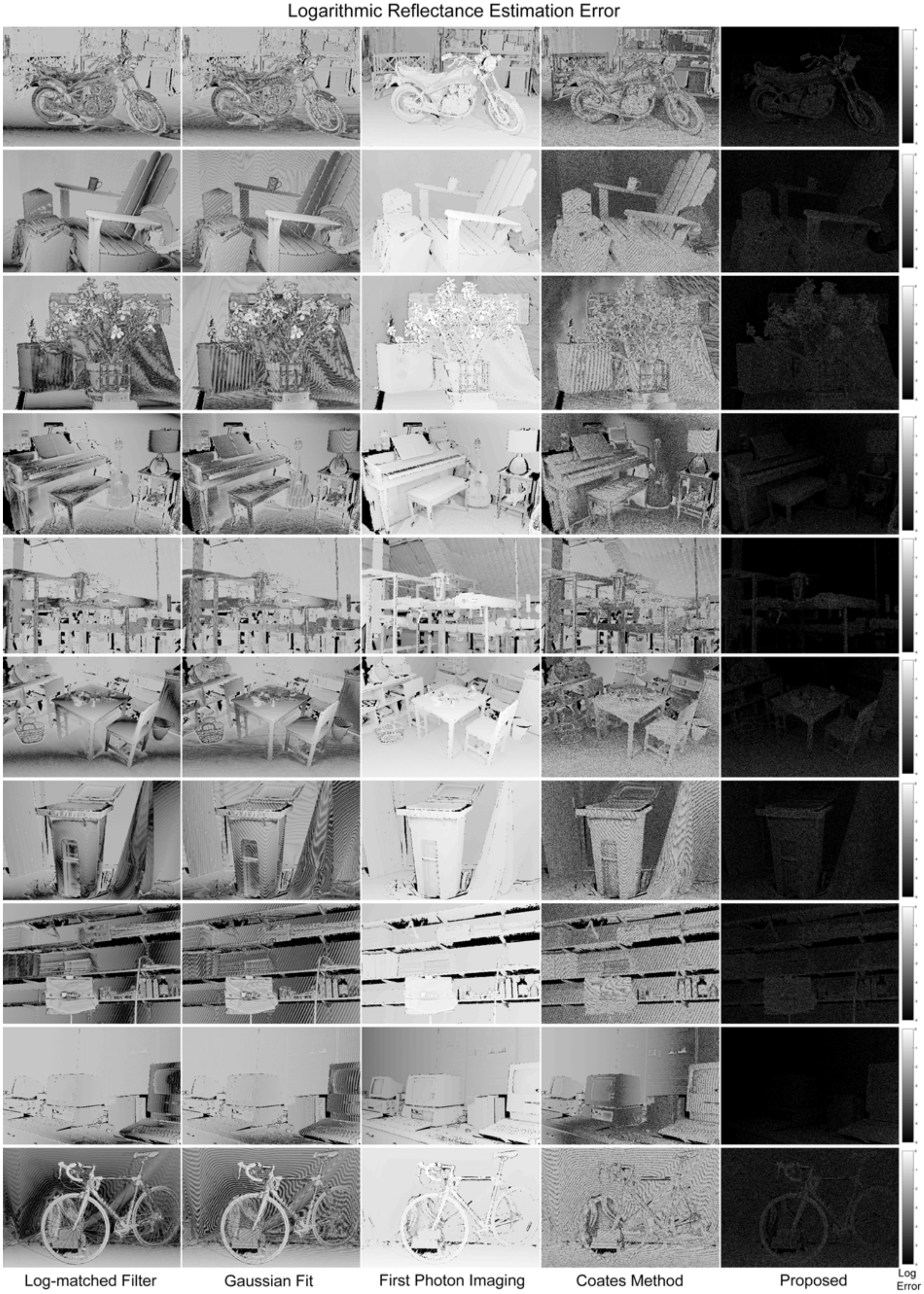

**Supplementary Figure 8:** Reflectance Error Maps an ideal Gaussian laser impulse response. Scenes from the Middlebury Stereo dataset [10] are shown.

**Supplementary Table 1:** Errors in depth and reflectance estimation for blue Alphalas LD-450-50 laser PSF (FWHM of 90 ps) are shown. Ground-truth depth and reflectance from each scene are taken from the Middlebury Stereo dataset [10]. The best depth and albedo result for each scene is highlighted. We compare here reconstructions for the conventional log-matched filter estimate [2], Coates method [3] followed by a Gaussian fit, first photon imaging [4]m and the proposed method

| Dataset Scene | Motorcycle | Adirondack | Jadeplant | Piano | Pipes | Playroom | Playtable | Recycle | Shelves | Vintage | Bicycle1 | **Average** |
|---|---|---|---|---|---|---|---|---|---|---|---|---|
| Log-matched Filter Avg. Depth Error [mm] | 13.29 | 15.44 | 15.08 | 12.81 | 10.56 | 14.07 | 14.40 | 14.57 | 14.66 | 10.90 | 17.61 | 13.95 mm |
| Coates [3] Avg. Depth Error [mm] | 2.96 | 3.73 | 3.83 | 2.58 | 1.85 | 3.27 | 3.50 | 3.43 | 4.16 | 1.71 | 5.45 | 3.32 mm |
| First Photon [4] Avg. Depth Error [mm] | 12.47 | 21.94 | 4.87 | 16.39 | 6.28 | 7.92 | 19.69 | 17.38 | 10.06 | 11.27 | 22.57 | 13.71 mm |
| Proposed Avg. Depth Error [mm] | **0.04** | **0.04** | **0.04** | **0.04** | **0.03** | **0.04** | **0.04** | **0.04** | **0.03** | **0.04** | **0.04** | **0.04 mm** |
| Log-matched Filter Reflectance PSNR [dB] | 11.28 | 9.73 | 9.53 | 13.06 | 15.24 | 10.33 | 10.43 | 11.13 | 9.35 | 16.10 | 7.42 | 11.24 dB |
| Coates [3] Reflectance PSNR [dB] | 17.08 | 15.59 | 15.21 | 18.55 | 19.94 | 15.79 | 16.41 | 17.02 | 15.04 | 20.53 | 13.40 | 16.78 dB |
| First Photon [4] Reflectance PSNR [dB] | 12.21 | 8.25 | 19.33 | 12.80 | 23.59 | 11.78 | 8.71 | 9.66 | 10.33 | 18.13 | 8.07 | 12.99 dB |
| Proposed Reflectance PSNR [dB] | **36.04** | **41.09** | **37.73** | **29.02** | **33.22** | **27.88** | **30.59** | **32.47** | **23.63** | **30.64** | **32.42** | **32.29 dB** |

**Supplementary Table 2:** Errors in depth and reflectance estimation for red Alphalas LD-670-50 laser PSF (FWHM of 50 ps) are shown. Ground-truth depth and reflectance from each scene are taken from the Middlebury Stereo dataset [10]. The best depth and albedo result for each scene is highlighted.

| Dataset Scene | Motorcycle | Adirondack | Jadeplant | Piano | Pipes | Playroom | Playtable | Recycle | Shelves | Vintage | Bicycle1 | **Average** |
|---|---|---|---|---|---|---|---|---|---|---|---|---|
| Log-matched Filter Avg. Depth Error [mm] | 5.34 | 6.31 | 6.16 | 5.21 | 4.14 | 5.74 | 5.87 | 6.00 | 6.08 | 4.28 | 7.26 | 5.67 mm |
| Coates [3] Avg. Depth Error [mm] | 2.35 | 2.82 | 2.76 | 2.05 | 1.28 | 2.50 | 2.68 | 2.70 | 3.19 | 1.48 | 4.01 | 2.53 mm |
| First Photon [4] Avg. Depth Error [mm] | 8.52 | 13.38 | 3.20 | 4.99 | 10.27 | 6.27 | 12.45 | 9.68 | 4.60 | 8.11 | 9.47 | 8.27 mm |
| Proposed Avg. Depth Error [mm] | **0.02** | **0.02** | **0.02** | **0.02** | **0.02** | **0.02** | **0.02** | **0.02** | **0.02** | **0.02** | **0.02** | **0.02 mm** |
| Log-matched Filter Reflectance PSNR [dB] | 12.22 | 12.80 | 12.00 | 15.45 | 11.70 | 12.92 | 12.88 | 15.20 | 12.97 | 12.72 | 10.53 | 12.85 dB |
| Coates [3] Reflectance PSNR [dB] | 8.41 | 6.48 | 7.50 | 6.70 | 9.73 | 6.81 | 7.06 | 5.87 | 6.25 | 9.74 | 6.86 | 7.40 dB |
| First Photon [4] Reflectance PSNR [dB] | 11.07 | 8.28 | 18.75 | 12.06 | 21.67 | 9.01 | 8.55 | 9.47 | 10.32 | 18.01 | 8.16 | 12.30 dB |
| Proposed Reflectance PSNR [dB] | **36.02** | **41.11** | **37.70** | **29.00** | **33.23** | **27.87** | **30.58** | **32.45** | **23.62** | **30.64** | **32.43** | **32.24 dB** |

**Supplementary Table 3:** Errors in depth and reflectance estimation for an ideal Gaussian impulse response (FWHM of 50 ps) are shown. Ground-truth depth and reflectance from each scene are taken from the Middlebury Stereo dataset [10]. The best depth and albedo result for each scene is highlighted.

| Dataset Scene | Motorcycle | Adirondack | Jadeplant | Piano | Pipes | Playroom | Playtable | Recycle | Shelves | Vintage | Bicycle1 | **Average** |
|---|---|---|---|---|---|---|---|---|---|---|---|---|
| Log-matched Filter Avg. Depth Error [mm] | 6.63 | 7.50 | 7.28 | 6.65 | 5.62 | 6.93 | 7.08 | 7.31 | 6.99 | 5.86 | 7.91 | 6.89 mm |
| Coates [3] Avg. Depth Error [mm] | 1.52 | 2.08 | 2.17 | 1.22 | 0.66 | 1.77 | 1.90 | 1.81 | 2.40 | 0.58 | 3.37 | 1.77 mm |
| First Photon [4] Avg. Depth Error [mm] | 2.59 | 7.28 | 2.94 | 2.79 | 3.58 | 5.58 | 5.10 | 6.16 | 4.27 | 2.67 | 7.03 | 4.54 mm |
| Proposed Avg. Depth Error [mm] | **0.02** | **0.02** | **0.02** | **0.02** | **0.02** | **0.02** | **0.02** | **0.02** | **0.02** | **0.02** | **0.02** | **0.02 mm** |
| Log-matched Filter Reflectance PSNR [dB] | 16.02 | 17.99 | 16.85 | 17.37 | 14.40 | 16.92 | 17.56 | 19.24 | 19.39 | 14.65 | 19.08 | 17.22 dB |
| Coates [3] Reflectance PSNR [dB] | 17.22 | 17.27 | 16.17 | 18.90 | 17.71 | 17.40 | 17.11 | 17.53 | 15.56 | 18.54 | 14.52 | 17.08 dB |
| First Photon [4] Reflectance PSNR [dB] | 12.56 | 8.66 | 19.38 | 15.22 | 23.80 | 9.43 | 9.42 | 10.46 | 10.48 | 19.52 | 8.59 | 13.41 dB |
| Proposed Reflectance PSNR [dB] | **35.82** | **40.49** | **37.29** | **28.99** | **33.19** | **27.85** | **30.52** | **32.35** | **23.61** | **30.62** | **32.24** | **32.09 dB** |

**Low-Flux Evaluation** The first-photon imaging method [4, 1] assumes a low-flux regime with negligible pileup and therefore performs only on-par with the conventional log-matched filter results in the quantitative evaluations from Tables 1-3. In addition to the pileup experiments from above, we have evaluated the proposed method in a low-flux regime at average photons $10^{-4}$ per-pulse and $N = 10,000$ shots. Table 4 shows depth and reflectance reconstruction performance in this scenario. Here, the first-photon imaging method achieves substantial improvements over the conventional log-matched filter method and Coates [3] pileup correction method. The latter performs on-par with the log-matched filter approach as measurement distortions due to pileup are minimal. The proposed method outperforms all competing approaches even in this low-flux scenario, however, it does not achieve sub-picosecond accuracy for such noisy measurement. Existing imaging methods tailored to this scenario, such as [2, 4, 1], perform denoising, deconvolution, peak-fitting, and histogram correction, as separate stages of a pipeline, for both reflectance and depth. In contrast, the proposed method performs these tasks jointly as part of the probabilistic inference framework from Supplementary Materials and Methods section, leading to the substantially improved performance even in this low-flux imaging scenario.

**Supplementary Table 4:** Low-Flux Evaluation. Errors in depth and reflectance estimation for red Alphalas LD-670-50 laser PSF (FWHM of 50 ps) are shown. Ground-truth depth and reflectance from each scene are taken from the Middlebury Stereo dataset [10]. The best depth and albedo result for each scene is highlighted.

| Dataset Scene | Motorcycle | Adirondack | Jadeplant | Piano | Pipes | Playroom | Playtable | Recycle | Shelves | Vintage | Bicycle1 | **Average** |
|---|---|---|---|---|---|---|---|---|---|---|---|---|
| Log-matched Filter Avg. Depth Error [mm] | 18.28 | 20.25 | 20.08 | 17.81 | 12.94 | 18.50 | 19.20 | 19.72 | 19.17 | 16.51 | 21.81 | 18.57 mm |
| Coates [3] Avg. Depth Error [mm] | 18.28 | 20.21 | 20.07 | 17.78 | 12.94 | 18.53 | 19.22 | 19.65 | 19.21 | 16.49 | 21.81 | 18.56 mm |
| First Photon [4] Depth Error [mm] | 6.87 | 5.57 | 8.63 | 8.16 | 11.51 | 4.34 | 3.16 | 8.09 | 9.01 | 1.89 | 12.55 | 7.25 mm |
| Proposed Avg. Depth Error [mm] | **0.89** | **0.65** | **0.67** | **0.73** | **1.11** | **0.65** | **0.71** | **0.65** | **0.54** | **1.12** | **0.55** | **0.75 mm** |
| Log-matched Filter Reflectance PSNR [dB] | 10.04 | 10.27 | 9.91 | 11.83 | 8.79 | 10.51 | 10.42 | 11.40 | 10.33 | 10.40 | 8.72 | 10.24 dB |
| Coates [3] Reflectance PSNR [dB] | 10.04 | 10.27 | 9.91 | 11.83 | 8.79 | 10.51 | 10.42 | 11.40 | 10.33 | 10.40 | 8.72 | 10.24 dB |
| First Photon [4] Reflectance PSNR [dB] | 18.09 | 18.75 | 18.67 | 20.65 | 19.99 | 18.67 | 19.61 | 19.55 | 19.68 | 23.54 | 17.90 | 19.55 dB |
| Proposed Reflectance PSNR [dB] | **26.89** | **31.25** | **26.99** | **25.86** | **28.35** | **23.98** | **27.05** | **29.39** | **21.32** | **27.88** | **26.36** | **26.85 dB** |

## Analysis

In this section, we analyze the effects of acquisition parameters on the reconstruction performance. To evaluate the probabilistic reconstruction method independent of the spatial priors, we consider a single-pixel histogram. Fig. 9 shows plots of the logarithmic depth estimation error with varying incident intensity, which is determined by the laser intensity, scene reflectance, and distance falloff. The effect of varying the number of shots $N$ per histogram acquisition is shown in individual plots for each of the two calibrated laser impulse response functions (cf. Fig. 2). The accuracy follows a similar trend across all plots. The conventional log-matched reconstruction reaches an optimum for a relatively low number of counts that is dropping quickly as the number of incident detections becomes larger than 1 expected photon detection per pulse, creating substantial histogram pileup. Existing sequential pileup correction methods, such as Coates [3], alleviate pileup and effectively extend the range where optimal performance can be reached beyond 1 expected photon detection per pulse. In this experiment, the proposed probabilistic reconstruction accuracy is identical to previous methods for the low-flux regime for a low number of shots since we only consider a single pixel and no spatial priors can be exploited. As the intensity increases the proposed approach substantially improves on all existing methods with the optimal performance by two orders of magnitudes for a higher number of shots. Note that even in low-flux regimes, the discussed method improves on competing approaches when pileup becomes negligible for a large number of shots. Also note that the optimal performance in each setting is achieved,

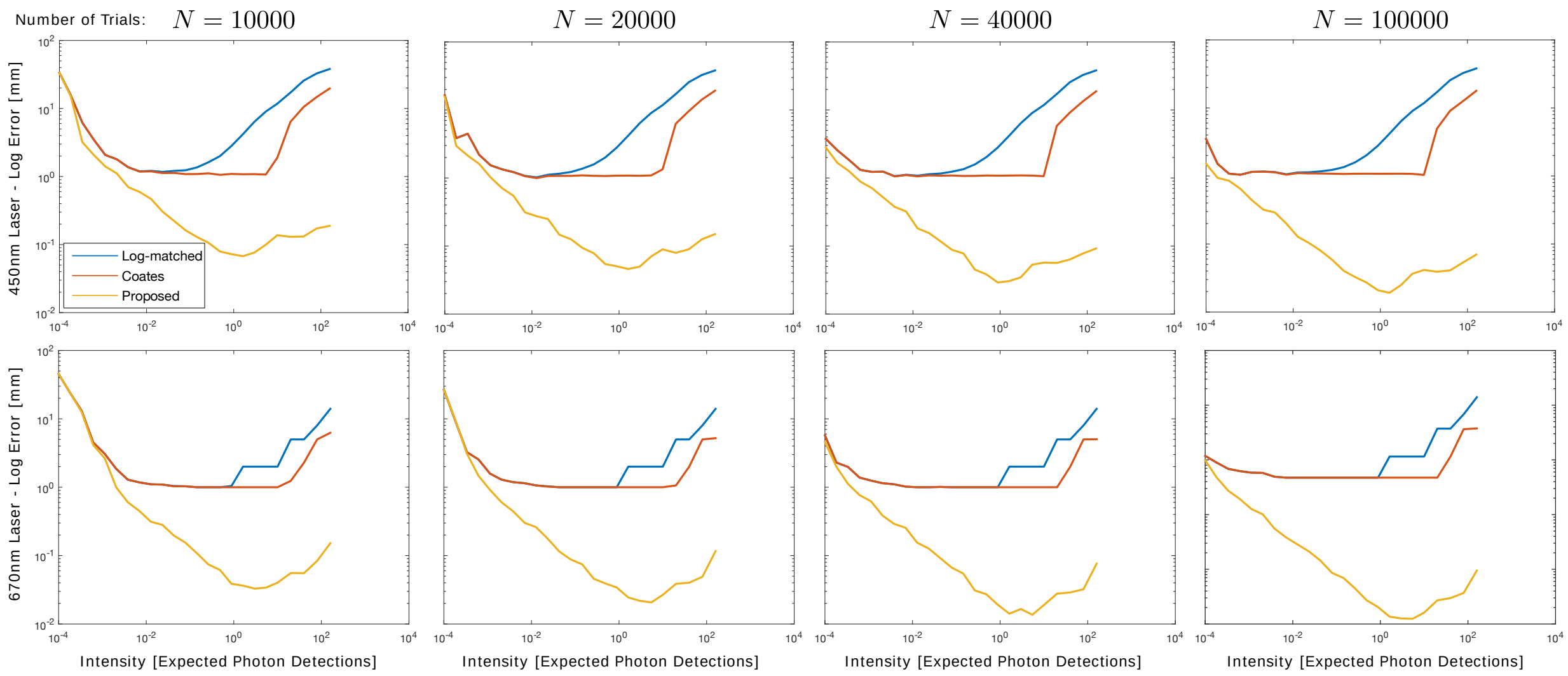

**Supplementary Figure 9:** We show the depth reconstruction accuracy for a varying number of shots $N$ at varying incident intensities which depend on distance of the reflector, reflectance and laser intensity. The first row depth accuracy results for the blue Alphalas LD-450-50 laser (FWHM of 90 ps) , and the second row shows the red Alphalas LD-670-50 laser (FWHM of 50 ps). Results are shown for the conventional log-matched filter method, Coates [3], and the proposed methods.

in fact, in a *different intensity region* around 1 photon detection per pulse, in contrast to existing methods which perform best between $10^{-2}$ to $10^{-1}$ average photons detections per pulse. Hence, the proposed approach argues not only for a joint probabilistic reconstruction

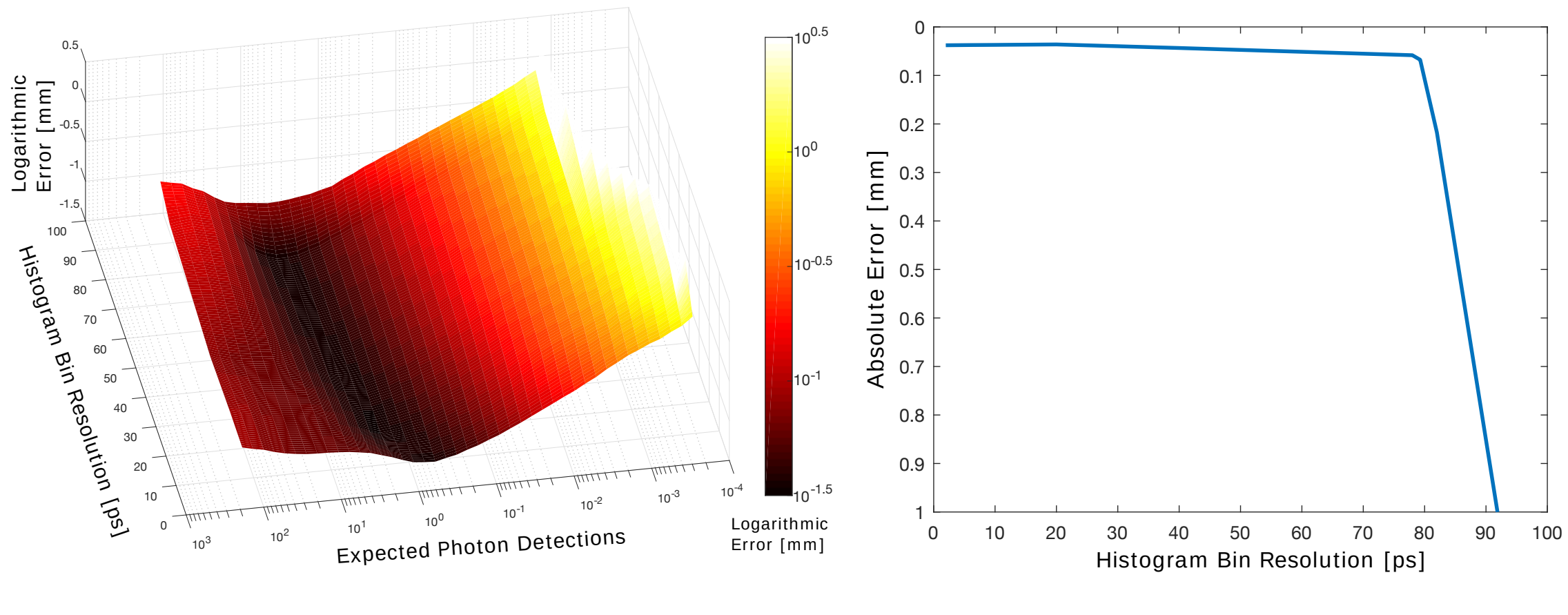

**Supplementary Figure 10:** Effect of Histogram Bin Width on Estimation Error. The left plot shows the depth reconstruction error of the proposed method for varying histogram bin sizes and incident intensities. We can see that the optimal intensity region stays around an average of 1 photon per pulse independent of the bin-size. The right plot shows reconstruction quality versus bin width for a fixed intensity of 1 photon per pulse. The proposed method achieves low depth reconstruction error across a broad range of bin widths, and only fails for bin widths larger than 80 ps.

but also for a different operating mode far outside of the conventional range.

Fig. 10 analyzes the effect of the histogram bin width on depth estimation accuracy in the same single-pixel scenario as considered above for Fig. 9. The accuracy of the proposed approach is consistent across a broad range of histogram bin-width from sub-2 picoseconds up to 80 picoseconds, as shown in the right plot of Fig. 10. This result verifies that the proposed method exploits the large temporal support of the laser impulse response. Only for very large bin-widths above 80 picoseconds, reconstruction accuracy begins to drop substantially. This means that the proposed approach, besides improving state-of-the-art accuracy by orders of magnitude, also *reduces the required timing accuracy* on the detector side. Moreover, the left plot in Fig. 10 verifies that the optimal intensity region lies around 1 photon per pulse independent of the histogram bin widths, demonstrating that the proposed method enables this unconventional operating regime across detectors with different temporal resolutions.

## Experimental Evaluation

In the following, we evaluate the proposed method on measurements experimentally acquired with the setup illustrated in Fig. 1 of the main manuscript. First, we validate the sub-picosecond accuracy of the proposed approach without using spatial priors, i.e. relying solely on the probabilistic pileup model for individual histogram measurements. Next, we assess the depth and albedo reconstruction quality on a variety of scenes with varying object shape and albedo. Both the per-histogram evaluations and the full spatio-temporal results demonstrate an order of magnitude improvement in accuracy compared to existing reconstruction methods.

### Temporal Resolution Evaluation without Spatial Priors

To validate the sub-picosecond accuracy of the described approach, we acquire a horizontal scanlines of histogram measurements. Three scene objects with different vertical depth profiles are used in the following experiments. Photographs of these scenes are shown in the left column of Fig. 11 along with an overlay illustration of the respective laser scanline path. All targets have been placed at a distance of 1 m to the detector. The first scene is a planar foam board target placed perpendicularly to the optical axis of the detector's objective lens. The second scene is a 3D-printed step target which was covered with a thin layer of white primer to eliminate severe subsurface scattering in the acrylic plastic printing material. With the third scene, shown in the bottom left of Fig. 11, we evaluate the performance of the proposed reconstruction method in the absence of spatial priors. Specifically, we acquire a sequence of single-spot measurements of a planar foam board target which is moved back and forth along the optical axis of the detector optics.

For all scenes, we acquire ground truth measurements by placing a 5% Neutral Density filter in the laser path which eliminates pileup distortion but leads to substantial increase in noise for a fixed number of shots. To eliminate these fluctuations in the ground truth reference, we acquire long sequences of 10 s length per spot at 5 MHz laser repetition rate. For these reference measurements, we use a histogram bin size of 4 ps. We calibrate the temporal impulse response from measurements of a planar target with the identical capture settings. The ground truth depth is then extracted from the clean, long-exposure measurements by log-matched filtering [2]. We acquire histogram measurements for each scene from $N = 100,000$ trials without additional filtering in the optical path. Comparing the best approaches from Depth and Reflectivity Error Evaluation section, we compare the proposed method against the conventional log-matched filter estimate [2], and Coates pileup correction method [3] followed by a Gaussian fit.

Figs. 11 and 12 show the ground truth and reconstructions for all scenes and for different laser impulse responses from the 450 nm Alphalas LD-450-50 laser and the 670 nm Alphalas LD-670-50 laser, see the Additional Calibration Details section for a comparison of the two temporal impulse response functions. We have removed the calibration offset due to the baseline between camera and sensor for the planar target for better visualization of the reconstruction error. The measurements show varying pileup based on laser spot position and response function due to the change in incident intensity based on distance-falloff and laser pulse energy. To illustrate the depth dependency, we have acquired the linear stage measurements at two different distances in Fig. 11. The measurement at 0.5 m distance suffers here from significantly more pileup distortion than the measurement at 1 m distance.

Table 5 shows the average absolute error in depth and round-trip time for all of these measurements. The proposed method outperforms all methods by more than an order of magnitude in improvement. In addition to the conventional log-matched filter estimate [2] and Coates method [3], we also show here Shin [4] applied on the pileup-corrected histogram data output by Coates method, which adds censoring and background signal suppression to the basic Coates method. The results validate that the proposed method achieves sub-picosecond accuracy with relatively long laser pulses larger than 50 ps independent of the laser pulse shape and independent of the scene shape. Note that the motion stage reconstructions have a higher average error than the planar target measurements due to the lack of spatial priors. This validates the performance independent of the scene (sub-picosecond accuracy is achieved for all scenes), but also demonstrates the performance of the spatial prior even in this high-intensity scenario, leading to an $2\times$ reduction in average error.

The depth profiles for the log-step in Figs. 11 and 12 reveal that the increased average error compared to the planar target occur at two positions where the target contains sharp edge transitions. At these positions, the depth of two surface patches at different depths are mixed in the impulse response, making this an ambiguous point with uncertainty in the log-matched ground-truth measurement. Note that the proposed approach solves here for a single depth estimate consistent with spatial prior information nearby. Aside from such corner-cases, the proposed method achieves low reconstruction error with sub-picosecond accuracy for all scanline measurements and translation measurements.

### Depth and Reflectivity Error Evaluation

In this section, we evaluate the performance of the proposed method for a variety of challenging real-world scenes with highly varying reflectance and depth profiles. Photographs of the test scenes are shown in Fig. 13. The individual scene objects contain complex geomet-ries, varying materials and reflectance properties, including highly specular and Lambertian reflectors with both spatially varying and uniform albedo. Hence, the intensity-driven pileup distortions occur with high spatial fluctuation for the given test scenes. We have acquired histogram measurements using the setup illustrated in Fig. 1 of the main manuscript using the 670 nm Alphalas LD-670-50 laser. The scene objects are placed at a distance of 1 m to the detector. For all scenes, we capture a ground truth reference measurement with a 5% Neutral Density filter in the laser path which eliminates pileup distortions by damping the source intensity. To minimize shot noise fluctuations at the low count rates resulting from the ND filter, we acquire very long sequences of 6 s length per spot at 4 MHz laser repetition rate. We scan every scene at a spatial resolution of $150 \times 150$ laser spots. Hence, a full reference measurement is acquired in $150 \cdot 150 \cdot 6$ s $= 37.5$ h per scene. The ground truth depth is extracted from this long-exposure measurements using log-matched filtering [2] with the impulse response calibrated using a planar target captured under the same acquisition settings, and hence unaffected by pileup. The target measurement, which serves as input for the proposed method, is acquired without any ND filter in the optical path. In addition, to validate the performance of the proposed method in the low-flux regime as in the Low-Flux Evaluation section, we acquire a third low-flux measurement using a 1% Neutral Density filter resulting in a very low per-spot count, e.g. less than 5 for Fig. 14.

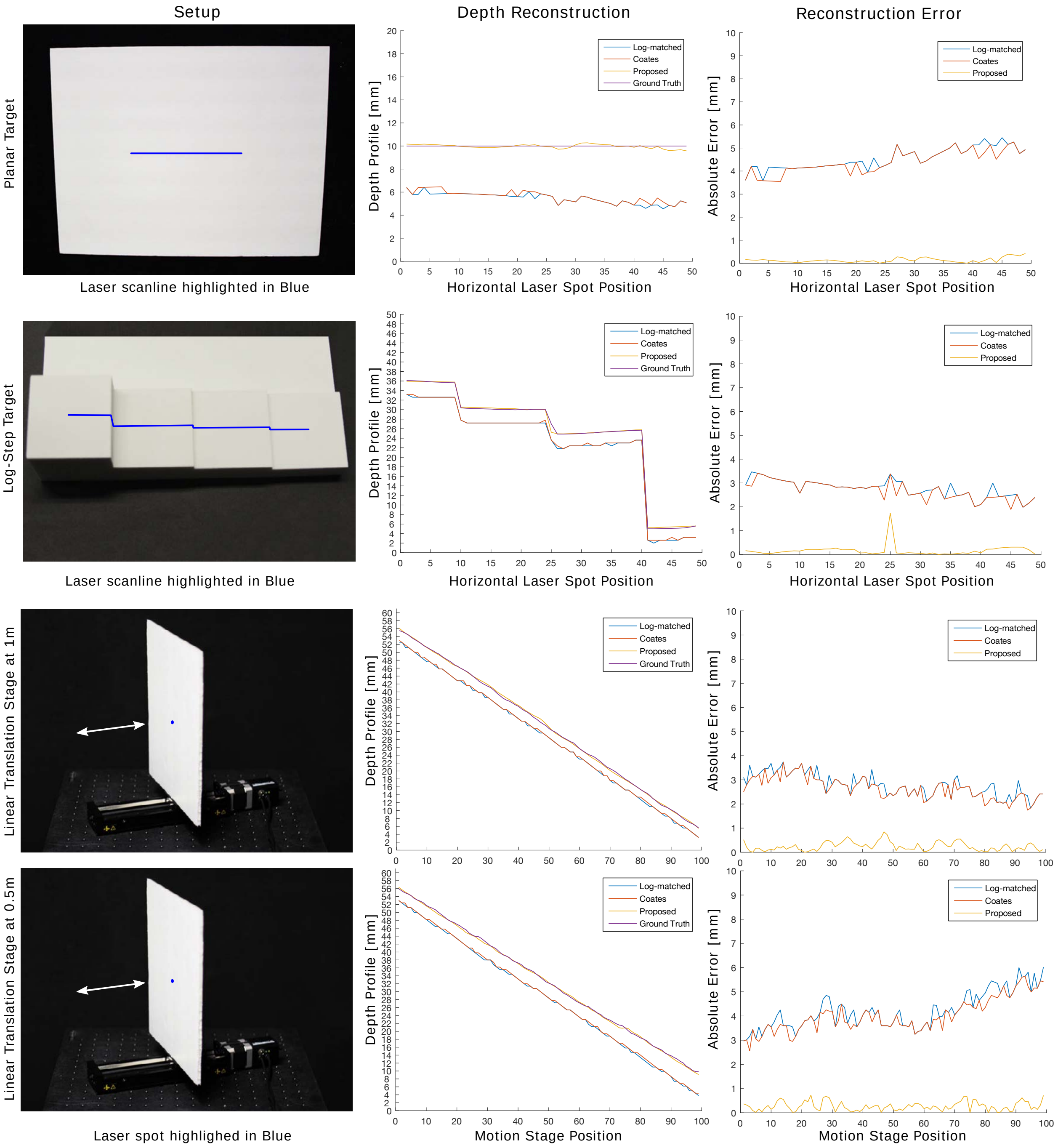

**Supplementary Figure 11:** Experimental depth evaluation for the 450 nm Alphalas LD-450-50 laser PSF (FWHM of 90 ps). Depth profiles for three different scenes with photographs in the left are shown in the rows. The first two rows show scanline measurements for a planar and step target, while the last two rows show a sequence of spot measurements for a motion stage aligned with the optical axis at the distance of 1 m and 0.5 m. The center column shows depth profile reconstructions for the conventional log-matched filter estimate [2], Coates method [3] followed by a Gaussian fit, and the proposed probabilistic reconstruction method. The right column shows the corresponding absolute error for each measurement. See Table. 5 for the corresponding error analysis.

**Supplementary Table 5:** Validation of sub-picosecond accuracy on experimental data. The average depth and round-trip time error for the scenes from Figs. 11 and 12 is shown, for the 450 nm Alphalas LD-450-50 laser (FWHM of 90 ps) and the 670 nm Alphalas LD-670-50 laser (FWHM of 50 ps), respectively. We compare here reconstructions for the conventional log-matched filter estimate [2], Coates method [3] followed by a Gaussian fit, Shin [4] on Coates-corrected measurements, and the proposed method.

| Scene Target<br>Laser Wavelength | Planar<br>450 nm | Log-Step<br>450 nm | Linear Stage<br>450 nm | Stage at 0.5 m<br>450 nm | Planar<br>670 nm | Log-Step<br>670 nm | Linear Stage<br>670 nm |
|---|---|---|---|---|---|---|---|
| Log-matched Filter Avg. Depth Error [mm] | 4.54 | 2.77 | 2.79 | 4.22 | 3.19 | 2.65 | 8.13 mm |
| Coates [3] Avg. Depth Error [mm] | 4.40 | 2.69 | 2.65 | 4.01 | 3.00 | 2.49 | 8.09 mm |
| First Photon [4] on Coates [3] Avg. Depth Error [mm] | 3.39 | 1.82 | 1.98 | 3.56 | 2.37 | 2.07 | 8.00 mm |
| Proposed Avg. Depth Error [mm] | **0.14** | **0.16** | **0.24** | **0.27** | **0.16** | **0.22** | **0.23 mm** |
| Log-matched Filter Avg. Round-time Error [ps] | 15.13 | 9.25 | 9.31 | 14.06 | 10.64 | 8.82 | 27.11 ps |
| Coates [3] Avg. Round-time Error [ps] | 14.68 | 8.96 | 8.85 | 13.35 | 9.99 | 8.29 | 26.97 ps |
| First Photon [4] on Coates [3] Round-time Error [mm] | 11.30 | 6.06 | 6.58 | 11.87 | 7.91 | 6.89 | 26.68 ps |
| Proposed Avg. Round-time Error [ps] | **0.46** | **0.55** | **0.80** | **0.91** | **0.52** | **0.75** | **0.76 ps** |

For each scene, depth and albedo reconstructions, and the corresponding error maps in the high-flux and low-flux regime are visualized in Figs. 14 - 18. Note that, in contrast to the previous section which validates the performance of the proposed method in the absence of spatial priors, we show here reconstructions using the proposed depth and reflectance priors. We show perspective renderings of the point cloud reconstructions along with error maps for the ground truth depth measured as discussed above. These results verify that the proposed approach achieves high-quality depth and albedo reconstructions for a wide range of scenes despite the substantial pileup distortion present in the measurements. The effect of pileup becomes apparent in the error map for the log-matched filter estimate [2] with more than 10 mm error depending on the scene. In contrast, the proposed method achieves sub-picosecond accuracy independent of scene depth and reflectance. Fig. 14 shows a scene with varying albedo and depth, Fig. 15 contains depth variations and uniform albedo, the scene in Fig. 16 is composed of cluttered scene geometry, Fig. 17 shows reconstructions for a highly specular copper coin surface, and Fig. 18 demonstrates results for the back side of the coin painted with matte primer. In all of these scenarios, the proposed approach recovers extremely fine features. The fine text on the bottom of the coin in Fig. 17 has a depth of less than 0.2 mm, measured with a coordinate measurement system, which is recovered by the proposed method. The depth profile of the Lincoln bust on the matte backside of the coin in Fig. 18 ranges between 0.05 mm and 0.3 mm and is faithfully recovered as visualized in the color-coded depth visualization in the center column of Fig. 18.

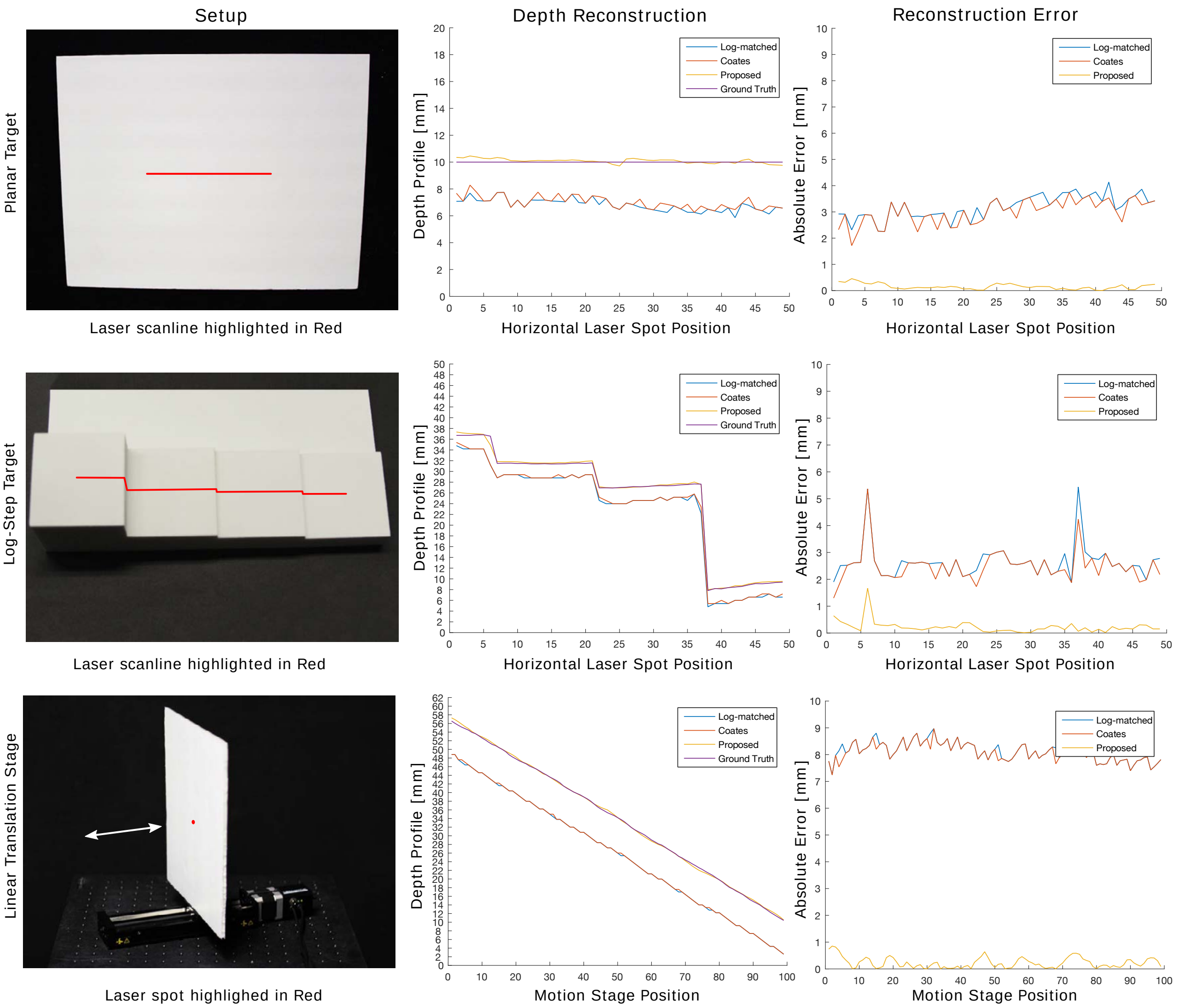

**Supplementary Figure 12:** Experimental depth evaluation for the 670 nm Alphalas LD-670-50 laser PSF (FWHM of 50 ps). The same scene as in Fig. 11 are shown here. See Table. 5 for the corresponding error analysis.

For all scenes, we have compared the proposed method against the best approaches identified in the Depth and Reflectivity Error Evaluation section, for both the high-flux and low-flux scenario. Specifically, we compare the proposed method against the conventional log-matched filter estimate [2] and Coates pileup correction method [3] followed by a Gaussian fit. The experimental error map results validate that the proposed method outperforms all existing methods by an order of magnitude in accuracy. We achieve sub-picosecond temporal resolution with relatively long laser pulses larger than 50 ps and independent of the scene. Note that this is in contrast to existing methods which do not account for pileup and hence suffer from strongly scene-dependent resolution.

In addition, we have also experimentally validated the proposed method in the low-flux regime in Figs. 14 - 18, previously evaluated in simulation in the Low-Flux Evaluation section. Specifically, we show error maps for the low-flux measurement for the first-photon imaging method [4] and the proposed method applied to the same low-flux measurement data. While the average error is substantially larger than in the high-flux range, matching the simulation results from the Analysis Section, the proposed method reduces the error by up to an order of magnitude for scene points, such as in Fig. 15. For areas with counts so low that the method only relies on the spatial prior, such as a few low-albedo regions in Fig. 14, the residual error is on par with the first photon imaging approach. However, across all test scenes, temporal resolution increases by a factor between 2× and 10× in the low-flux regime. While existing imaging methods tailored to this scenario, including [4], perform denoising, deconvolution, peak-fitting, and histogram correction, as separate stages of a pipeline, these experimental results validate that jointly solving these tasks in the proposed probabilistic inference framework improves resolution by an order of magnitude in both, the high-flux as well as low-flux measurement scenario.

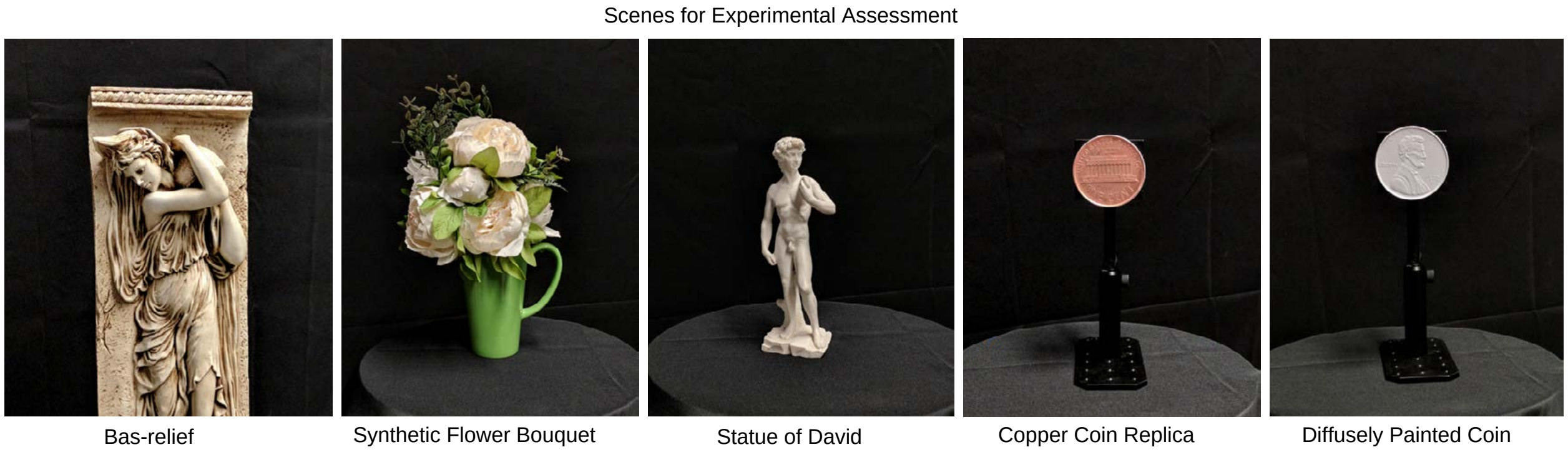

**Supplementary Figure 13:** Photographs of the scenes used in the experimental validation of the proposed approach. The scenes contain objects with complex shapes, e.g. synthetic flower scene, varying reflectance properties, e.g. metallic coin with highly specular reflectance, and varying albedos, e.g. painted bas-relief.

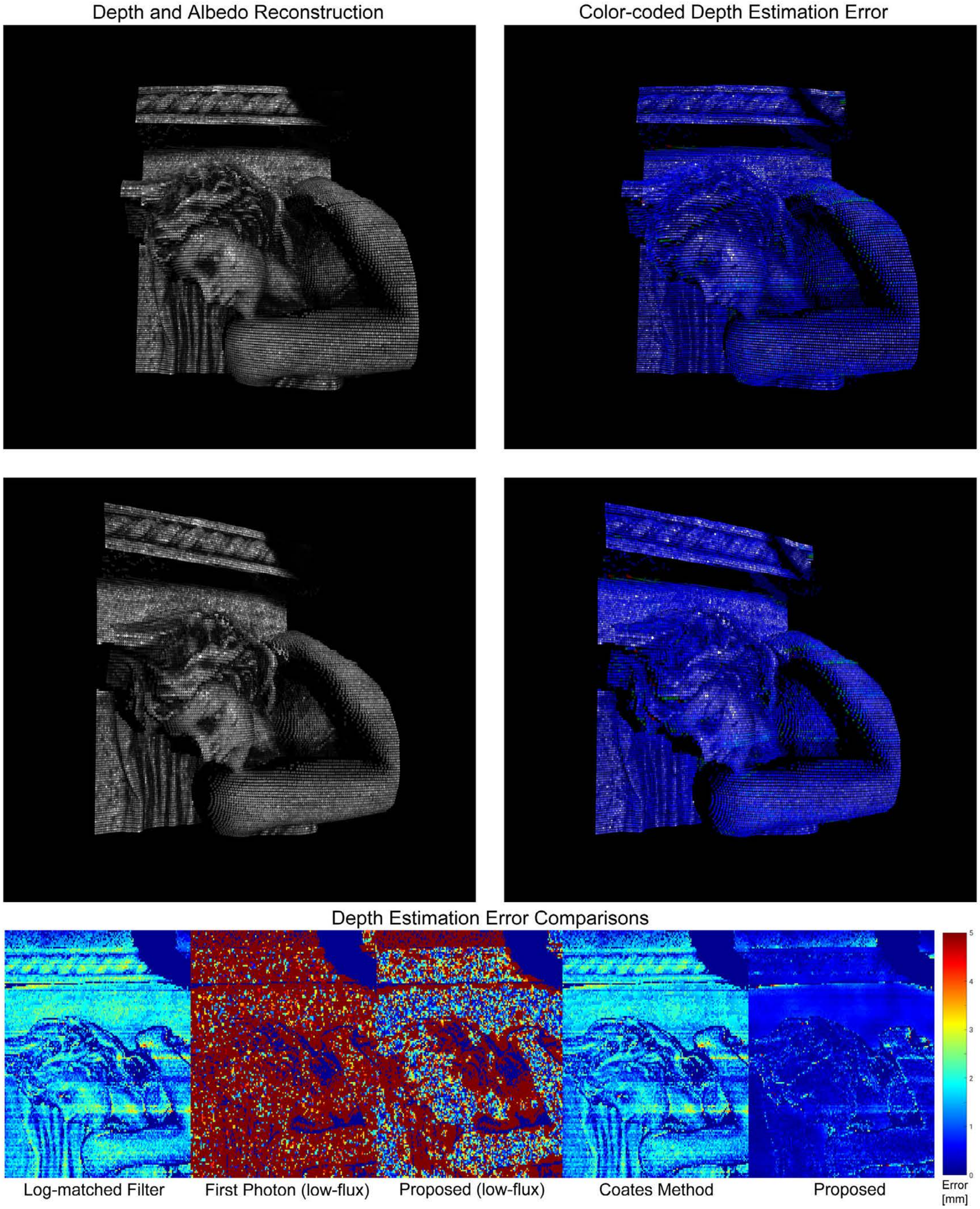

**Supplementary Figure 14:** Experimental results and error analysis for the "Bas-relief" scene from Fig. 13. Perspective point cloud renderings of depth and albedo are shown for two different camera positions in the top two rows. While the recovered reflectance is shown as point-intensity in the left plots, the right plots additionally color-code the error according to the color map in the bottom right. The bottom row visualizes the corresponding orthographic depth error maps comparing: conventional log-matched filter [2], first-photon imaging [4, 1] on a low-flux measurement of the scene, the proposed method on identical low-flux measurement, Coates method [3] followed by Gaussian fit (on high-flux measurement), and the proposed method.

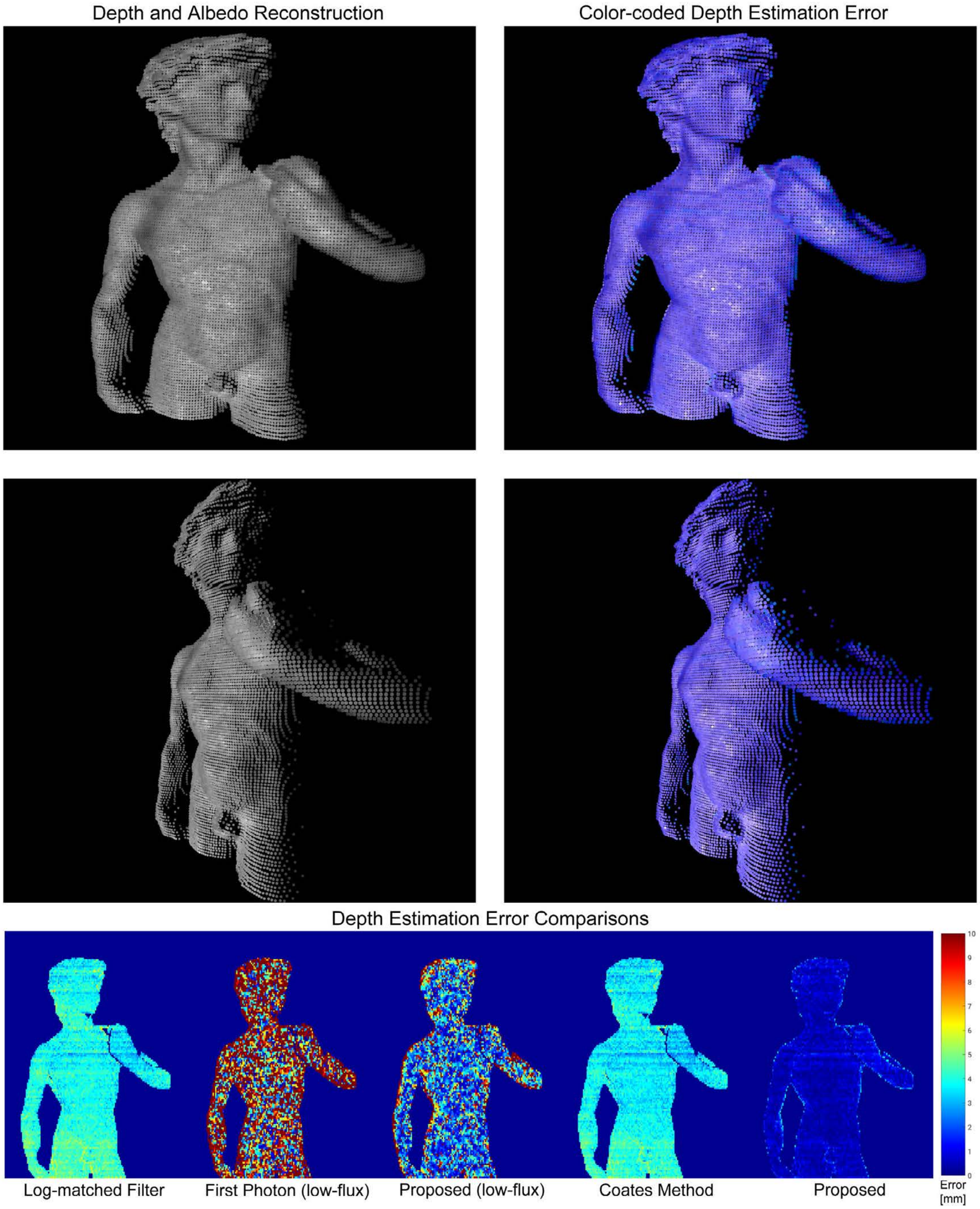

**Supplementary Figure 15:** Experimental results and error analysis for the "Statue of David" scene from Fig. 13. Perspective point cloud renderings and orthographic error maps are shown analogue to Fig. 14.

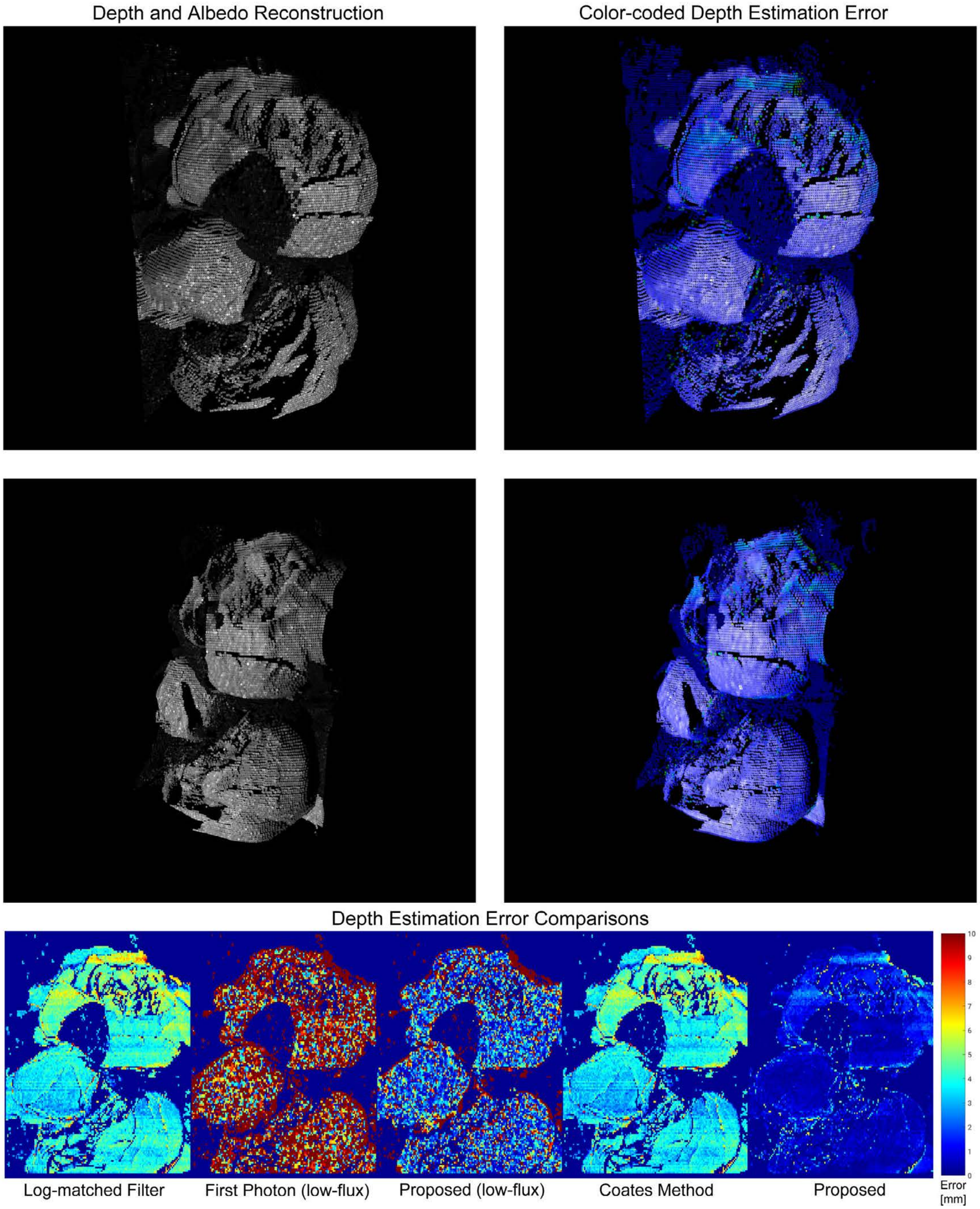

**Supplementary Figure 16:** Experimental results and error analysis for the "Synthetic Flower Bouquet" scene from Fig. 13. Perspective point cloud renderings and orthographic error maps are shown analogue to Fig. 14.

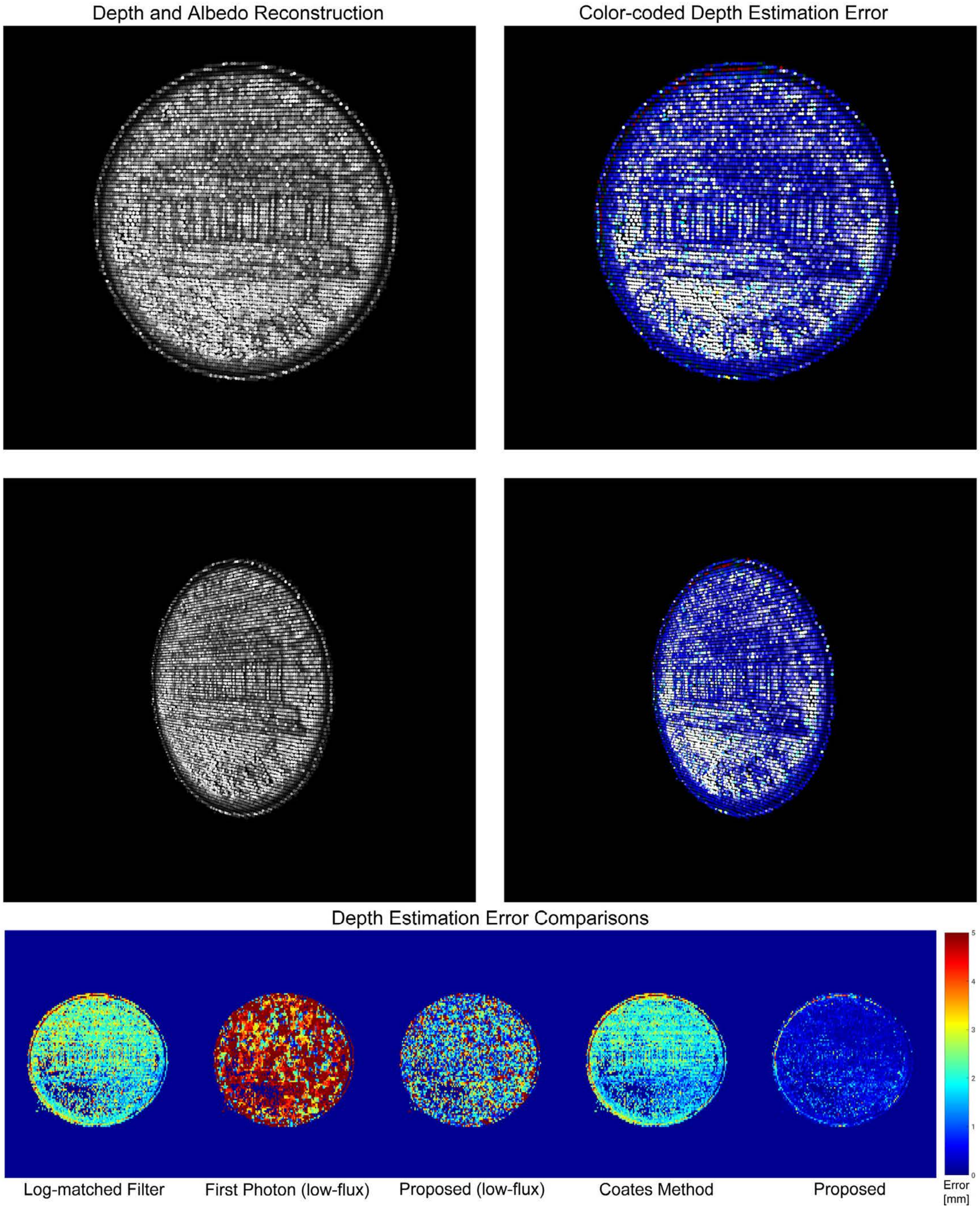

**Supplementary Figure 17:** Experimental results and error analysis for the “Copper Coin Replica” scene from Fig. 13. Perspective point cloud renderings and orthographic error maps are shown analogue to Fig. 14.

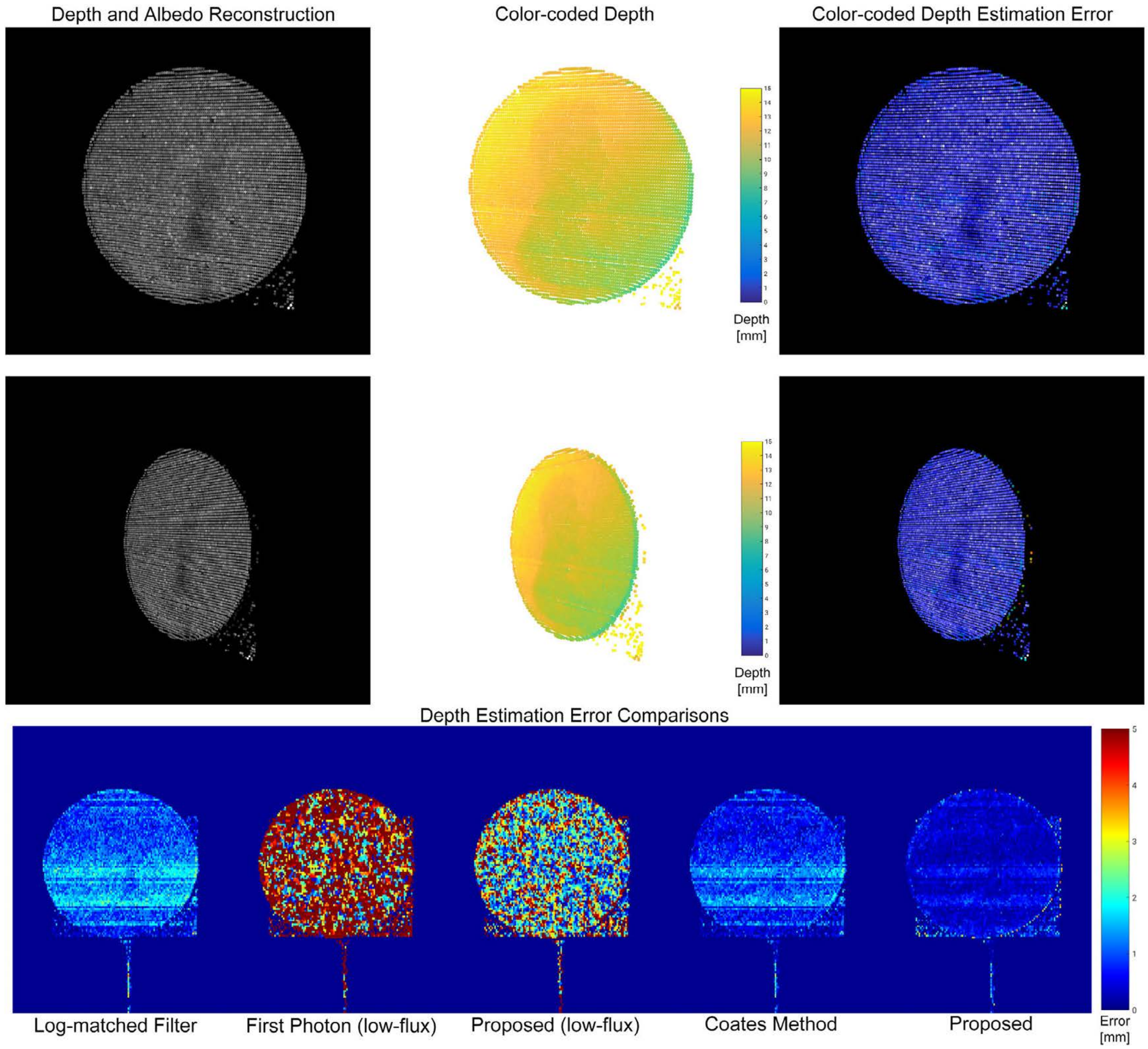

**Supplementary Figure 18:** Experimental results and error analysis for the "Diffusely Painted Coin" scene from Fig. 13. Perspective point cloud renderings and orthographic error maps are shown analogue to Fig. 14. In addition, point cloud renderings with the depth color-coded are shown in the center column of the top two rows, which highlight the fine depth profile of the Lincoln bust profile recovered in the center of the coin.